\documentclass[journal,twoside,web]{ieeecolor}
\usepackage{generic}
\usepackage{cite}
\usepackage{amsmath,amssymb,amsfonts}
\usepackage{graphicx}
\usepackage{textcomp}
\usepackage{url}
\usepackage{hyperref}

\usepackage{array}
\usepackage{algorithmicx}
\usepackage{algpseudocode}
\usepackage{algorithm}

\def\d{\mathrm{d}}

\def\u{\mathrm{u}}
\def\l{\mathrm{l}}
\def\b{\mathrm{b}}
\def\e{\mathrm{e}}
\def\m{\mathrm{m}}

\def\org{\mathrm{org}}
\newtheorem{problem}{Problem}

\def\BibTeX{{\rm B\kern-.05em{\sc i\kern-.025em b}\kern-.08em
		T\kern-.1667em\lower.7ex\hbox{E}\kern-.125emX}}
\begin{document}
	\title{Comprehensive Energy Footprint Benchmarking Algorithm for Electrified Powertrains}
	\author{Hamza Anwar
		%\IEEEmembership{Student Member, IEEE}
		, Aashrith Vishwanath, Apurva Chunodkar, and Qadeer Ahmed
		%\IEEEmembership{Member, IEEE}
		%\thanks{This research did not receive any specific grant from funding agencies in the public, commercial, or not-for-profit sectors.}
		\thanks{H. Anwar is a PhD student at the Center for Automotive Research, Ohio State University, Columbus, OH 43212 (e-mail: anwar.24@osu.edu).}
		\thanks{A. Vishwanath is a MS student at the Center for Automotive Research, Ohio State University, Columbus, OH 43212 (e-mail: vishwanath.16@osu.edu).}
		\thanks{A. Chunodkar is a Senior Technical Specialist - Controls \& Diagnostics Research at Cummins Inc., Columbus, IN 47201 (e-mail: apurva.chunodkar@cummins.com).}
		\thanks{Q. Ahmed is a Research Associate Professor at the Department of Mechanical \& Aerospace Engineering, Ohio State University, Columbus, OH 43212 (e-mail: ahmed.358@osu.edu).}}
	
	\maketitle
	\begin{abstract}
		Autonomy and electrification in automotive control systems have made modern-day powertrains one of the most complex cyber-physical systems. This paper presents a benchmark algorithm to quantify the performance of complex automotive systems exhibiting mechanical, electrical, and thermal interactions at various time-scales. Traditionally Dynamic Programming has been used for benchmarking the performance, however, it fails to deliver results for system with higher number of states and control lever due to curse of dimensionality.  
		%Due to a range of subsystems in hybrid electric vehicles exhibiting mechanical, electrical, and thermal interactions at various time-scales, optimal energy management to meet high-level system objectives is a challenging control problem. Recent research trends are going towards large-scale optimization of the powertrain with all its subsystems involving numerous state variables and control levers simultaneously. To this end, numerical optimization approaches have proven to be robust in handling the complex powertrain interactions. 
		We propose ``PS3'', a three-step algorithm for mixed-integer nonlinear optimal control problems with application to powertrain energy management. PS3 uses pseudo-spectral collocation theory for highly accurate modeling of dynamics. %In existing literature, none of the other state-of-the-art approaches 
		Based on the validated powertrain component models, we have addressed simultaneous optimization of electrical (SOC), vehicular (eco-driving) and thermal (after-treatment and battery temperatures) dynamics along with an integer (gear and engine on/off) control and its corresponding (dwell-time) constraints. 
		PS3 is used to solve such large-scale powertrain problems having fast and slow dynamic states, discontinuous behaviors, non-differentiable and linearly interpolated 1-D and 2-D maps, as well as combinatorial constraints. Five case study powertrain control problems are given to benchmark the accuracy and computational effort against Dynamic Programming. Our analysis shows that this algorithm does not scale computational burden as Dynamic Programming does, and can handle highly complex interactions that occur in modern-day powertrains, without compromising nonlinear and complex plant modeling. 
		%This methodology can be applied to difficult problem classes and objective functions having causal or non-causal controllers and having full or short-horizon settings.
	\end{abstract}
	
	
	\section{Introduction}
	\label{sec:intro}
	\IEEEPARstart{I}{n} modern times hybrid electric vehicles (HEVs) have increasingly become complex systems. Optimal energy management strategies (EMSs) consider the various subsystems of a powertrain, as well as their interactions to achieve targets of fuel economy along with emissions of air pollutants and greenhouse gases. Many a times, the objectives in an EMS will conflict with each other, such as minimizing the fuel and improving driveability performance. On the other hand, the subsystems of a powertrain may exhibit very different behaviors in their time-dynamics and control. For example, an electric battery may have rapid charging and discharging while on the other hand, an after-treatment catalyst may have slow increase in its temperature. Likewise, the transmission and clutch subsystems will have discontinuous shifts and switches, while the ratio of power-split between internal combustion engine and electric machine, could be any real-number within its bounds. Complicating it further, can be a scenario when in the same energy management problem eco-driving is allowed where one is not restricted to operate on a given drive cycle but is allowed to modulate the speed profile around a target profile.
	
\begin{figure}[!t]
		\centering
		\includegraphics*[width=\linewidth]{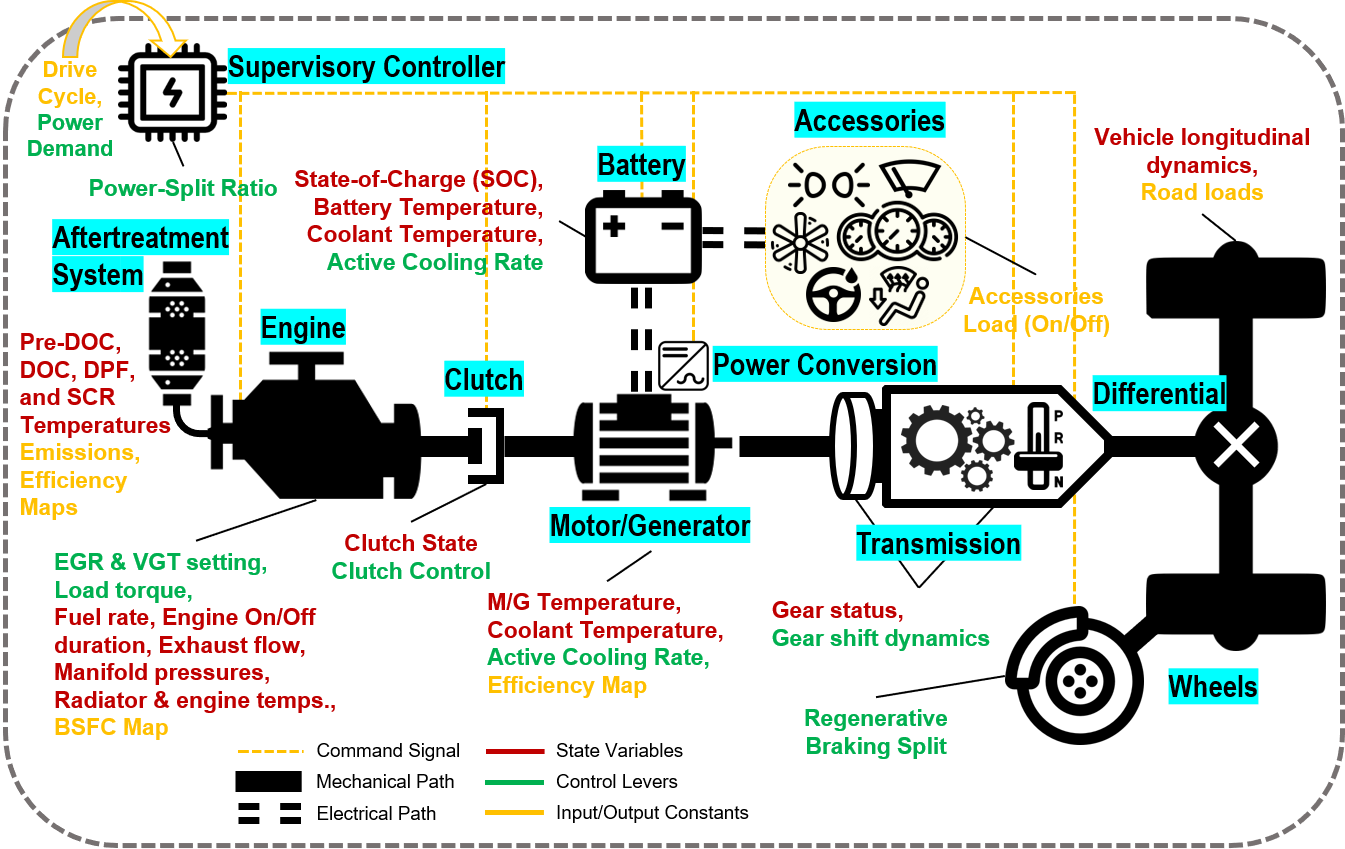}
		\caption{An example hybrid electrified powertrain with parallel architecture showing various subsystems with frequently used control and state variables for energy management.\label{fig:architecture}}
	\end{figure}

	For the energy management, these dynamic and discontinuous interactions restrict engineers to only model incomplete and approximated relationships or to model the state and control variables in disjoint sub-problems. And so traditional control strategies focus on individual sub-component optimality instead of a joint holistic system-level optimization. For example, the engine may be programmed to solely operate at its optimal operating line and the after treatment system independently tuned to maintain certain temperatures of its catalysts. But, whether or not all these systems will jointly meet the overall objectives of the powertrain operation remains a question. If supervisory controller is involved, then all system state variables and control levers may not be modeled and solved jointly with supervisory controller to completely capture such interactions. Hence, there is a great need to develop energy management approaches that can handle high degree of complex system-level interactions, tackle stiffness caused by fast and slow dynamics, and exhibit discontinuous and combinatorial interactions on the optimal control problem, while meeting the conflicting objectives of hybrid electric powertrains. An example of a powertrain with various subsystems, states and controls that are usually considered across powertrain control literature is shown in Fig.~\ref{fig:architecture}.
	
	\subsection{Related Work}
	\label{subsec:relatedwork}
	Research on energy management strategies for hybrid electric vehicles has a history of more than twenty years. The field has evolved and prospered to a great extent because of the ever-increasing applications of electric powertrains in the world \cite{tran2020thorough}. Various methods have been used to address powertrain control problems which can be broadly classified into rule-based and optimization-based approaches. Optimization-based approaches either tend to be focused on short-horizon and real-time capability, such as the ones using model-predictive-control (MPC) framework \cite{huang2017model,burger2019design} or ECMS \cite{wu2018optimized,onori2016hybrid}, or on full-horizon offline study. Any given optimization approach can further either be direct (first-discretize-then-optimize) or be indirect (first-optimize-then-discretize) in its methodology. 
	
	Among the most widely used offline EMSs is discrete Dynamic Programming (DP) which is a direct approach. Due to its popularity guaranteeing globally optimal solutions, its variants \cite{pan2014probabilistic,li2008dynamic,johannesson2007assessing} have found space in online EMSs as well. But, among the most common problems associated with it are the requirement of \textit{apriori} route information and its inherent curse of dimensionality which restricts it being scaled up to work for problems involving two or more continuous state variables. As for DP's variants, the critical step is to approximate the cost-to-go but such approximations are usually data-driven and specific to given drive cycles \cite{larsson2014cubic,powell2007approximate,lin2004stochastic}. If integer-valued variables (also called discrete variables) are optimized using DP, such as gear choices and engine on/off switch, it outperforms many of its competitor methods. Nonetheless, DP is considered a benchmark in literature especially for offline validation of new methods.
	
	Other important works have involved indirect optimization methods such as Pontryagin's Minimum Principle \cite{serrao2011comparative}. In particular, researchers have used PMP for eco-driving problems as well, wherein the power demand from the vehicle level is made flexible and is jointly optimized with the powertrain management controls like powersplit \cite{ghasemi2018powertrain}. However, indirect approaches, like PMP, have a drawback of requiring highly accurate Jacobian and Hessian forms of the variables involved. Another drawback is that they have difficulty in incorporating path constraints into the control problem, making them mathematical formulations intractable or impractical.
	
	Another classification of optimization based approaches is usually done into gradient-based optimization, like Newton's method and gradient-descent, versus derivative-free optimization. In HEV EMS, derivative-free optimization is not very-well developed, more so because of its independence from the physics of the problems, such as particle swarm optimization \cite{chen2016particle}. On the other hand, gradient-based optimization methods have long been used for energy management of HEVs in numerous ways. 
	
	\subsubsection*{Gradient-based Optimization for Energy Management in HEVs}
	In simple words, gradient-based approaches rely on finding minima of an objective function by determining the solution to its gradient subject to certain dynamic and algebraic constraints. Earliest of works that used gradient-based optimization for HEV energy management used linear programming \cite{pisu2003lmi}. Even recent works involve linear programming such as fast iterative-LP \cite{robuschi2020minimum} which used one real-valued state variable. The main consideration for LP based problems is to simplify and approximate the model into a linear form, such as the Willan's line approximation for the internal combustion engine.
	
	Moving on, gradient-based EMSs using convex analysis have also been a hot research topic \cite{marinkov2016convex}. These approaches rely on having convex models for the various subsystems and quadratic programming (QP) or semi-definite programming (SDP) concepts \cite{koot2005energy,reinbold2013global} are further used to obtain solutions using off-the-shelf solvers \cite{grant2014cvx}. More recently, the trend has shifted towards fast sequential-QP, and SDP-based integer convex minimization \cite{egardt2014electromobility,hadj2016convex,park2018semidefinite}. The limitations of convexity-based optimization is the difficulty to specify the highly nonlinear powertrain models as convex models. Secondly, discontinuities introduced by discrete variables or look-up tables in the model are difficult to handle if the framework restricts the form of the model to be convex.
	
	The approaches that make use of nonlinear programming (NLP) tools and solvers are not very old for HEV EMS. NLPs are difficult to solve as they do not assume convex or linear forms of the models, and instead they deal with system non-linearities as they are. To formulate the NLPs, there has been a trend in EMS context to utilize the theory of pseudo-spectral collocation (PSC) lately. PSC allows for highly accurate modeling of state dynamics, at the expense of NLP size. But with increasingly powerful computational resources nowadays, handling hundreds of differential equations accurately is possible through combining state-of-art NLP solvers, such as IPOPT \cite{wachter2006implementation}, WORHP \cite{buskens2012esa}, etc. PSC scheme has been employed frequently for various optimization problems in chemical industrial processes and aerospace research, but its adoption for automotive research sector is quite recent. Some works that have used PSC for formulating of NLPs and solved an HEV energy management control problem include \cite{zhou2015pseudospectral,wei2017pseudospectral,liu2019multi}.
	
	NLP-based optimization approaches can handle complex optimal powertrain control problems, but cannot easily handle discontinuities. In the context of HEVs, there are very few works that involve mixed-integer nonlinear programming (MINLP), such as \cite{robuschi2021multiphase}, which combines integer optimization employing combinatorial integral approximation (CIA) theory along with PSC. Being able to solve for continuous and discrete variables simultaneously by solving an MINLP is quite challenging. There are open-source \cite{bonami2008algorithmic} and commercial \cite{byrd2006k} MINLP solvers available as well in the market, but, the experts of MINLP field suggest that a homegrown solver specific to the problem being addressed is more useful and much faster compared to an off-the-shelf solver. An introductory tutorial to numerical optimization and mixed-integer programming can be found in the lecture notes \cite{diehl2016lecture} and the book \cite{sager2005numerical}. Concepts about nonlinear programming, its sparse solvers, its relationship with optimal control problems and similar important background can be found in \cite{betts2010practical}.
	
	For the purpose of this paper, we focus on developing and validating an HEV energy management algorithm involving large-scale optimization having high number of state and control variables. Our algorithm is based on MINLP employing PSC for transcription of the original mixed-integer optimal control problem into the MINLP. It has a three-step approach wherein we solve a relaxed version of the large MINLP in first step, solve for the integer-variables using mixed-integer quadratic programming (MIQP) in the second step, and after fixing integer variables we re-optimize the real-valued NLP variables in the third step. We show results on five case-study fuel-minimization problems with varying problem size and complexity to benchmark the computational-effort and optimality with Dynamic Programming solutions. In the sequel paper \cite{paper2citation}, we demonstrate and present detailed analysis of results for a thirteen-state and four-control variable problem where the objective function has conflicting terms of fuel and NOx emission given different relative weightings.
	
	\textbf{Paper structure:} Section \ref{sec:problem} describes the general formulation of powertrain control problems that we aim to solve. Section \ref{sec:mainsec} is about the proposed algorithm for solving any given powertrain control problem, while Section \ref{sec:results} shows numerical results on five case-study problems along with comparison to DP, and lastly conclusions are summarized in the Conclusion section.
	\section{Problem Description}
	\label{sec:problem}
	As described earlier, we aim to solve such optimal powertrain control problems which exhibit large number of dynamical states, combination of real and integer-valued controls, and constraints from various interacting powertrain components through complex relationships. This section gives details about the problem formulation structure that we consider for powertrain energy management problems we aim to solve. Through this section, we also explain the difficulties associated in modern-day HEV energy management problems when considering large-scale joint optimization.
	
	%Starting of with a known drive cycle that serves as reference for the vehicle, the objective function can be to minimize fuel, or emissions, or combination of both. Some examples of control signals that can be jointly solved using this framework are eco-driven vehicle acceleration, powersplit factor between engine and battery, gear selection, engine on/off switching. Likewise, the state dynamics and time-varying constraints can be of very different time-scales in their origin, such as thermal (e.g. after-treatment subsystem), electrical (e.g. battery), mechanical (e.g. electric machine and transmission), and chemical (e.g. engine combustion). This is an offline study with the potential of real-time capability subject to availability of look-ahead information sources.
	
	\subsection{Dynamic States and Controls}
	The generic optimal control problem can have continuous (real-valued) or discrete (integer-valued) state variables, denoted as $\mathbf{x}(t)\in \mathcal{X}\subset \mathbb{R}^{|\mathcal{X}|}$ and $\mathbf{x}_\d(t)\in \mathcal{X}_\d\subset \mathbb{Z}^{|\mathcal{X}_\d|}$ respectively, at time $t$. Likewise, there can be continuous control variables, $\mathbf{u}(t)\in \mathcal{U}\subset \mathbb{R}^{|\mathcal{U}|}$ and discrete control variables $\mathbf{u}_\d(t)\in \mathcal{U}_\d\subset \mathbb{Z}^{|\mathcal{U}_\d|}$. The sets $\mathcal{X},\mathcal{U},\mathcal{X}_\d,$ and $\mathcal{U}_\d$ are simply specified by constant lower and upper bounds on each of its variables, and are handled by \textit{box} constraints. The dynamics for the continuous state variables are specified by ordinary differential equations (ODE), whose right hand sides are nonlinear (and possibly discontinuous) functions of the state and control variables, as well as other dependent signals. 
	%\begin{gather*}\begin{bmatrix}
	%\mathbf{\dot{x}}(t)\\\mathbf{\dot{x}}_\d(t)\end{bmatrix} = \begin{bmatrix}\mathbf{f}(\mathbf{x}(t),\mathbf{u}(t),\mathbf{x}_\d(t),\mathbf{u}_\d(t),t)\\ \mathbf{f}_\d(\mathbf{x}_\d(t),\mathbf{u}_\d(t),t)\end{bmatrix}.
	%\end{gather*}
	As an example, we consider fuel consumption to be a state variable in our experiments which is given by linear interpolation of 2-D engine map, and hence has a discontinuous RHS. Dynamics of discrete state variables, on the other hand, are specified by their discrete-time dynamic equations. Examples of these are gear shifting and engine on/off switching. The reader is referred to the sequel \cite{paper2citation} where discrete dynamics are exclusively described and detailed.
	
	When properly initialized, discrete state variables are a consequence of discrete controls having dynamics dependent only on their respective discrete controls. For example, engine on/off switch is a discrete control taking values 0 (no change), 1 (turn on) and $-$1 (turn off). The engine status is a state variable with dynamics dependent only on engine on/off switch. Thus, its differential equation can be written as linear combination of shifted and scaled Dirac Delta functions, having impulses at engine on/off switch events. Consequently, the engine status variable will only take discrete values.
	
	The continuous state or control variables are classified into consistent state or control variables, $(\mathbf{x}_{\mathrm{con}}(t),\mathbf{u}_{\mathrm{con}}(t))$, and inconsistent state or control variables, $(\mathbf{x}_{\mathrm{inc}}(t),\mathbf{u}_{\mathrm{inc}}(t))$. Consistent variables are those which can be defined in a way that their dynamics depend neither on inconsistent nor discrete state or control variables. On the other hand, the inconsistent variables can have dependence on any state or control variable. Thus, we have the three types of state ODEs:
	\begin{gather}\label{eq:stateDyn}\begin{bmatrix}
			\mathbf{\dot{x}}_\mathrm{con}(t)\\\mathbf{\dot{x}}_\mathrm{inc}(t)\\\mathbf{\dot{x}}_\d(t)\end{bmatrix} = \begin{bmatrix}\mathbf{f}_\mathrm{con}(\mathbf{x}_\mathrm{con}(t),\mathbf{u}_\mathrm{con}(t),t)\\
			\mathbf{f}_\mathrm{inc}(\mathbf{x}(t),\mathbf{u}(t),\mathbf{x}_\d(t),\mathbf{u}_\d(t),t)\\
			\mathbf{f}_\d(\mathbf{x}_\d(t),\mathbf{u}_\d(t),t)\end{bmatrix},
	\end{gather}
	where, $\mathbf{{x}}(t)= \begin{bmatrix}
		\mathbf{{x}}_\mathrm{con}(t) & \mathbf{{x}}_\mathrm{inc}(t)\end{bmatrix}^\top$, $\mathbf{{u}}(t)= \begin{bmatrix}
		\mathbf{{u}}_\mathrm{con}(t) & \mathbf{{u}}_\mathrm{inc}(t)\end{bmatrix}^\top$, and the vector-valued functions $\mathbf{f}_{(\cdot)}$ are general form expressions of the RHSs of respective ODEs which solely depend on the way system dynamics are modeled. Distinction of continuous state variables into consistent and inconsistent naturally arises due to the separation caused with discrete variables, $\mathbf{{x}}_\mathrm{d}(t)$ and $\mathbf{u}_\mathrm{d}(t)$. For example, in a backward powertrain model when gear status is considered to be a discrete state and the vehicle speed as a continuous state, then the vehicle speed, distance and acceleration are all consistent variables since system causality dictates that these vehicle-level variables come from a known drive cycle, and gear selection depends on those but it will not be the other way around. In that case, all powertrain variables after transmission will be classified as inconsistent variables because they all have dependence on gear selection. In general, wherever there is a discrete variable, we try to split the overall system there into the two continuous variable types. In our case studies, this classifies vehicle-level variables (speed, distance and acceleration) as consistent and powertrain-level variables (SOC, torque split, after-treatment temperatures, etc.) as inconsistent. However, note that if such separation of (\ref{eq:stateDyn}) is not possible for some problem, then all continuous variables will be considered as inconsistent variables.
	%For different phases of the drive-cycle, we may have different set of dynamics for the states, which is why the problem is multi-phase. By different phases, we mean traction, braking and stand-still vehicle operation which is indicated by time periods specified by $\tau_{\mathrm{total}}(t)\geq 0$, $\tau_{\mathrm{total}}(t)<0$ and $v(t)=0$, respectively, where $\tau_{\mathrm{total}}$ is the demand torque after transmission.
	\subsection{Boundary Constraints}
	All state variables require initial conditions to be defined. These are known constants defining values of every state variable at the initial time. Along with initial conditions for the states, some state variables also have constraint on the final-value they take. For the problems presented in the results section, we impose charge-sustaining constraint on battery's state of charge. For problems considering eco-driving, the total distance covered by the eco-driving vehicle at speed $v(t)$ must be the same as the total distance covered by the reference speed profile $v_{\org}(t)$.
	\begin{gather}\label{eq:boundCon}\begin{bmatrix}
			\mathbf{x}(t)\\\mathbf{x}_\d(t)\end{bmatrix}_{t=0} = \begin{bmatrix}\mathbf{x}_0\\ \mathbf{x}_{\d,0}\end{bmatrix},\qquad \begin{bmatrix}
			\mathbf{x}(t)\\\mathbf{x}_\d(t)\end{bmatrix}_{t=T} = \begin{bmatrix}\mathbf{x}_T\\ \mathbf{x}_{\d,T}\end{bmatrix}.
	\end{gather}
	\subsection{Algebraic Relationships}
	In order to characterize the plant behavior completely, we define another set of variables which are functions of each other as well as of the states and controls. These are coupled in the optimal control problem through algebraic relationships. We term them \textit{signals}. For example, during traction phase in parallel HEV, the electric machine torque, $\tau_\m$ and total demand torque after transmission, $\tau_\mathrm{total}$ are related to torque split, $\mu$ through: $\tau_\m = \mu \tau_\mathrm{total}$. The total demand torque is, likewise, algebraically related to the vehicle speed $v$ and acceleration $a$ through transmission efficiency, gear number $g$, and various road load signals. Another example is of the fuel consumption engine-out NOx emissions which are based on 2-D look-up tables of engine shaft speed $\omega$ and engine torque $\tau_{\e}$. If any of these \textit{signals} is constrained, then that is effectively a type of \textit{path} constraint on states or controls. These relationships are numerous ranging from kinematic equations at vehicle and driveline level, to energy conservation and efficiency losses in between propulsion (engine and traction motor), after-treatment and driveline blocks, as well as thermal heat transfer and electric current dynamics. For our case study problems, these relationships and dependent variables are listed in relevant modeling sections of sequel \cite{paper2citation}.
	\subsection{Path and Box Constraints}
	Each of the state and control variables is bounded below and above by known constants, called box constraints. All box constraints are specified as vectors by subscripts $(\cdot)_{\l\b}$ and $(\cdot)_{\u\b}$ for lower and upper bounds, respectively:
	\begin{align}\label{eq:boxCon}\begin{split}
			\mathbf{u}_{\l\b} \leq \mathbf{u}(t)& \leq \mathbf{u}_{\u\b}\\
			\mathbf{x}_{\l\b} \leq \mathbf{x}(t)& \leq \mathbf{x}_{\u\b}\\
			\mathbf{u}_{\d,\l\b} \leq \mathbf{u}_\d(t)& \leq \mathbf{u}_{\d,\u\b}\\
			\mathbf{x}_{\d,\l\b} \leq \mathbf{x}_\d(t)& \leq \mathbf{x}_{\d,\u\b}
	\end{split}\end{align}
	
	Along with box constraints, the state and control variables can have explicit or implicit constraints that are time-varying. These are jointly termed as path constraints. Examples of explicit path constraints on a state variable are the eco-driving speed constraint and stop-at-stop constraint | vehicle speed $v(t)$ is constrained to be within a constant envelope of $V_{\mathrm{margin}}=5$ km/h above and below the reference speed profile, $v_{\org}(t)$ and should stop when there is a stop in the reference:
	\begin{align*}
		|v_{\org}(t) - v(t)| \leq \begin{cases}
			V_{\mathrm{margin}}&\textrm{if } v_{\org}(t)\neq 0\\
			0&\textrm{if } v_{\org}(t)= 0
		\end{cases}
	\end{align*}
	Likewise, example of implicit path constraints can be the time-varying min/max limits on \textit{signals} such as engine or motor torques.
	\begin{gather*}
		\tau_{\e,\min}(t) \leq \tau_{\e}(t)\leq \tau_{\e,\max}(t)\\
		\tau_{\m,\min}(t) \leq \tau_{\m}(t) \leq \tau_{\m,\max}(t)
	\end{gather*}
	One important path constraint is the dwell-time constraint on a discrete variable. For example, if the controller optimizes gear profile to minimize fuel consumption, we will observe gear chattering phenomenon. But it is undesirable for gears to rapidly switch here and there as that causes immense drivability discomfort. Hence an explicit path constraint is needed on gear switching that limits number of gear shifts for a certain dwell-time period $t_\mathrm{dwell}$. This is a combinatorial constraint on a discrete state variable and we give details about how we handle this in case-study problem of the sequel paper \cite{paper2citation}.
	%\[\mathbf{0} \leq \mathbf{\Gamma}_{\mathrm{dwell}}(\mathbf{x}_\d(t_{-}),t_{-})\quad\forall t_{-}\in [t\!-\!L,t]\]
	Path constraints can be grouped as:
	\begin{gather}\label{eq:pathCon}\mathbf{h}\left(\mathbf{x}(t),\mathbf{u}(t),\mathbf{x}_\d(t),\mathbf{u}_\d(t),t\right)\leq 0\end{gather}
	\subsection{The Optimal Control Problem}
	\label{subsec:ocp}
	Finally, using (\ref{eq:stateDyn})(\ref{eq:boundCon})(\ref{eq:boxCon})(\ref{eq:pathCon}) we arrive at the complete mixed-integer optimal control problem, Prob.~\ref{prb:ocp}, where, $\mathbf{{x}}(t)= \begin{bmatrix}
		\mathbf{{x}}_\mathrm{con}(t) & \mathbf{{x}}_\mathrm{inc}(t)\end{bmatrix}^\top$, and $\mathbf{{u}}(t)= \begin{bmatrix}
		\mathbf{{u}}_\mathrm{con}(t) & \mathbf{{u}}_\mathrm{inc}(t)\end{bmatrix}^\top$. The cost function comprises of a running cost $L$ and a terminal cost $\psi$. For numerical results in this paper we have considered total fuel consumption
	% as shown in Prob.~\ref{prb:ocp}, but it can be replaced by any combination of the state variables, control levers, or the plant signals used in the model
	over the whole cycle as the cost function. Note that, in the following definition, we have particularly identified the vectors in boldface, time-varying signals with ``$(t)$'' and constants without ``$(t)$''.
	% As is customary practice, the ``$(t)$'' indicates the signal value at time $t$.
	\begin{problem}[$\mathbb{OCP}\mathnormal{1}$: Optimal Powertrain Control Problem]\label{prb:ocp}
		\begin{align*}
			\min_{{\mathbf{u}(\!t\!),}{\mathbf{u}_\d(\!t\!)}}\quad
			&\psi(\mathbf{x}(T),\mathbf{x}_\d(T),\!T)\!+\!\!\int_{0}^{T}\!\!\!\!\!L(\mathbf{x}(t),\mathbf{u}(t),\mathbf{x}_\d(t),\mathbf{u}_\d(t),\!t)\d t
			\\
			\mathrm{s.t.}\quad
			&\textit{ODEs:}
			\begin{cases}
				\mathbf{\dot{x}}_\mathrm{con}(t)= \mathbf{f}_\mathrm{con}(\mathbf{x}_\mathrm{con}(t),\mathbf{u}_\mathrm{con}(t),t)\\
				\mathbf{\dot{x}}_\mathrm{inc}(t)=\mathbf{f}_\mathrm{inc}(\mathbf{x}(t),\mathbf{u}(t),\mathbf{x}_\d(t),\mathbf{u}_\d(t),t)\\
				\mathbf{\dot{x}}_\d(t)=\mathbf{f}_\d(\mathbf{x}_\d(t),\mathbf{u}_\d(t),t)
			\end{cases}
			\\
			&\textit{Box Constr.:}
			\begin{cases}
				\mathbf{u}_{\l\b} \leq \mathbf{u}(t) \leq \mathbf{u}_{\u\b}\\
				\mathbf{x}_{\l\b} \leq \mathbf{x}(t) \leq \mathbf{x}_{\u\b}\\
				\mathbf{u}_{\d,\l\b} \leq \mathbf{u}_\d(t) \leq \mathbf{u}_{\d,\u\b}\\
				\mathbf{x}_{\d,\l\b} \leq \mathbf{x}_\d(t) \leq \mathbf{x}_{\d,\u\b}
			\end{cases}
			\\
			&\textit{Path Constr.:}
			\begin{cases}
				\mathbf{h}\left(\mathbf{x}(t),\mathbf{u}(t),\mathbf{x}_\d(t),\mathbf{u}_\d(t),t\right)\leq 0
			\end{cases}
			\\
			&\textit{Boundary Constr.:}
			\begin{cases}
				\mathbf{x}(0)=\mathbf{x}_0\\
				\mathbf{x}_\d(0) = \mathbf{x}_{\d,0}\\
				\mathbf{x}(T)=\mathbf{x}_T\\
				\mathbf{x}_\d(T) = \mathbf{x}_{\d,T}
				%\\\mathbf{h}_T\left(\mathbf{x}(T),\mathbf{x}_\d(T),T\right)\leq 0
			\end{cases}
		\end{align*}
	\end{problem}
	
	\section{PS3: A Three-Step Algorithm for HEV EMS}
	\label{sec:mainsec}
\begin{figure*}[!t]
		\centering
		\includegraphics*[width=\textwidth]{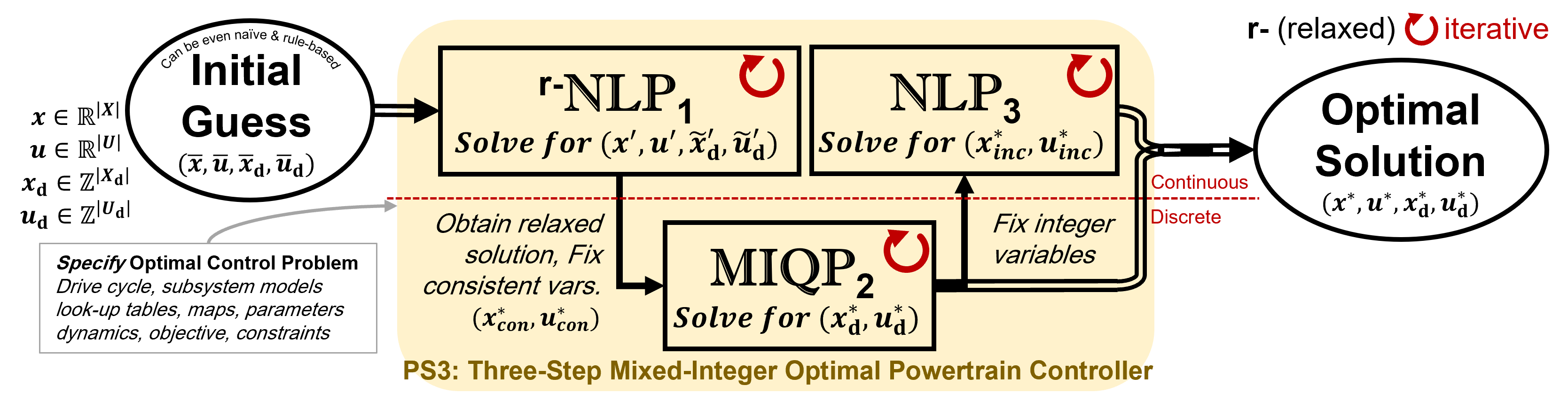}
		\caption{Proposed three-step algorithm ``PS3'' for mixed-integer optimal powertrain control.\label{fig:algo_diag}}
	\end{figure*}

	A direct method of numerical optimization is used to solve the optimal control problem. Direct methods rely on a first-discretize-then-optimize approach. All numerical optimization methods at some point rely on an iterative approach towards finding solutions. Insofar, the underlying principle approach is to iteratively progress in the gradient direction such that a minima is found within specified tolerance levels. As described earlier, we make use of a customized MINLP solution approach, which is done in three steps, solving $\mathbb{NLP}\mathnormal{1}$, $\mathbb{MIQP}\mathnormal{2}$ and $\mathbb{NLP}\mathnormal{3}$ in each step respectively, depicted in Fig.~\ref{fig:algo_diag}. In the following, we will explain the three steps in detail.
	\subsection{Discretization using Pseudo-Spectral Collocation}
	\label{subsec:psc}
	Discretization of the optimal control problem is the first crucial step in solving it using a numerical optimization technique. Furthermore, some of the constraints in our problem definition, that are related to gear dwell-time, can be better described only after an equivalent discrete-time problem is defined. Hence, before we attempt to solve the optimal control problem to determine a solution, we discretize it in time. Once an equivalent discrete-time optimization problem is defined, we can then move on to formulating our three-step approach to solve the resultant mixed-integer nonlinear program (MINLP). For this discretization of the continuous-time optimal control problem into a discrete-time numerical optimization problem we make use of the \textit{pseudo-spectral collocation} theory.
	
	The pseudo-spectral method is essentially a high-order implicit Runge-Kutta (IRK) based collocation scheme in which the time-axis is discretized at non-uniform locations which are determined based on roots of a certain family of orthogonal polynomials. These polynomials are employed to accurately approximate the state trajectories originating from the differential equations that govern the plant dynamics in optimal control problems. Due to high accuracy of derivatives and integrals that comes via such an approximation, pseudo-spectral collocation has gained a lot of popularity.
	
\begin{figure}[!t]
		\centering
		\includegraphics*[width=\linewidth]{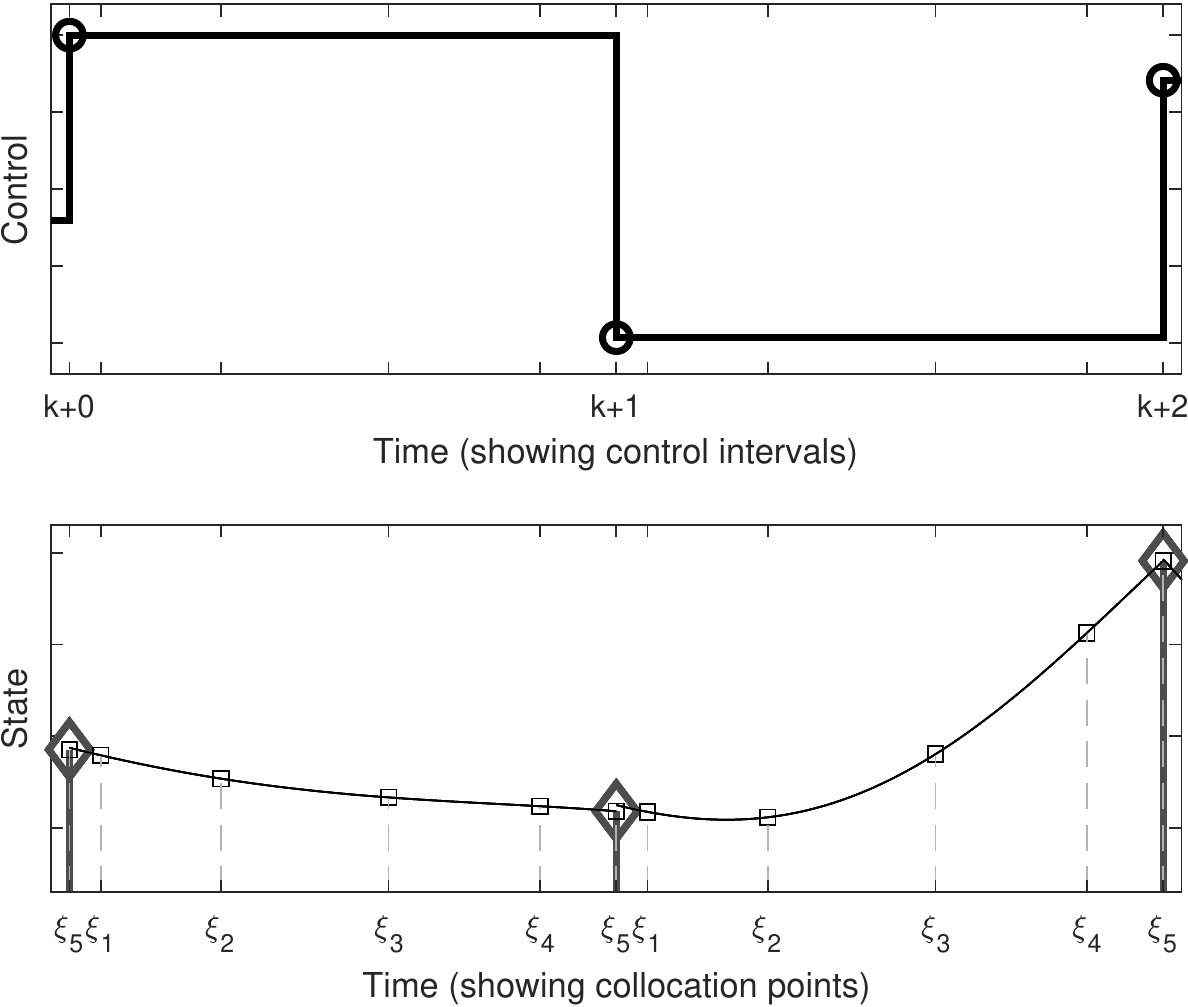}
		\caption{Discretization of continuous time OCP with 5 Radau collocation points ($\xi_i \in \{0.06,0.28,0.58,0.86,1\}$ on an interval $(0,1]$ for $i=1,\cdots,5$) in each control interval to ensure state continuity and smoothness. Control signal is piece-wise constant in each interval of 1 second duration, and corresponding state signal is determined by a 5-th order polynomial using Radau scheme\label{fig:finiteelement}}
	\end{figure}

	We use a discretization step size of, $\Delta t := \Delta t_k = 1$ {second} $\forall k\in \{1,\cdots,N\}$ for the $k$-th time interval indicated by the time $t\in [t_{k-1},t_k)$, and having a total of $N$ such intervals, called `control intervals', spanning the complete time horizon $[0,T]$. For notational convenience, when dealing with discrete-time signals we use $k$ in parentheses instead of $t$. The control signals are assumed to be piece-wise constant within each of the $N$ intervals.
	%Within $k$-th control interval out of total $N$ equally spaced intervals, the control signals are assumed to be piece-wise constant:
	%\begin{align*}\mathbf{u}(t) =\mathbf{u}(k)~&\forall~{t\in [t_{k-1},t_k)},\\\mathbf{u}_\d(t) = \mathbf{u}_\d(k)~&\forall~{t\in [t_{k-1},t_k)}.\end{align*}
	On the other hand, the state trajectories smoothly vary due to high-order implicit Runge-Kutta (IRK) discretization at collocation points. For highly accurate state dynamics modeling we use five LGR (Legendre-Gauss-Radau) collocation points within each control interval
	%, $\sigma_{k}^i\in (t_{k-1},t_{k}]$ for $i \in \{1,2,3,4,5\}$, within $k$-th control interval
	| see Fig.~\ref{fig:finiteelement}. But, practically, for some of our experiments when accuracy is not expected to be compromised or when a benchmark needs to be compared, we use one collocation point per interval.
	%Thus, state trajectories are approximated using collocation scheme and denoted at $i$-th LGR point in the $k$-th control interval of the $N$ total intervals as:
	%\begin{align*}\mathbf{x}^i(k) := \mathbf{x}(t)~&\mid~{t\in [t_{k-1},t_k)} \land \sigma_{k}^i\in (t_{k-1},t_{k}].\end{align*}
	One specialty of LGR collocation scheme, unlike other choices of LGL (Legendre-Gauss-Lobotto) or LG (Legendre-Gauss) schemes, is that it includes the interval's end point as a collocation point. Thus, LGR collocation points have stiff decay property that can well handle stiffness associated with the corresponding ODEs. We omit details of how the pseudo-spectral collocation scheme operates to achieve discretization at non-uniform points inside a control interval, and refer the reader to exclusive works on the subject by \cite{rao2009survey} and \cite{biegler2010nonlinear}.
	%By doing such discretization, we transcribe our OCP into a mixed-integer nonlinear program expressed in a symbolic modeling language, which can be iteratively optimized using an off-the-shelf nonlinear programming solver.
	%\[\sigma_1~\sigma_2~\sigma_3~\sigma_4~\sigma_5\]
	%\[t_{i-3}~t_{i-2}~t_{i-1}~t_{i}~t_{i+1}~t_{i+2}~t_{i+3}\]
	\subsection{Constructing the relaxed-Nonlinear Program}
	\label{subsec:nlp}
	Since the original optimal control problem (Prob.~\ref{prb:ocp}) involves discrete state and control variables, discretization of the same using pseudo-spectral collocation into $N$ control intervals, as mentioned above, will transcribe it into a mixed-integer nonlinear program. Instead of numerically solving the MINLP directly, we first apply relaxation to its discrete variables that allows the solver to assume continuous values for the otherwise discrete-valued variables, $(\mathbf{x}_\d(k),\mathbf{u}_\d(k))~\forall k\in \{1,\cdots,N\}$. Thus, if we only have one discrete state, the gear number, $g(t)$, then the \textit{relaxed} gear number state, $\tilde{g}(t)$ discretized into $N$ intervals, $\tilde{g}(k)~\forall k\in \{1,\cdots,N\}$ can be any real number within $1=:g_{\l\b} \leq \tilde{g}(k)\leq g_{\u\b} :=n_{\mathrm{b}}$, where $n_{\mathrm{b}}$ is the total number of gear choices here. For a six-speed transmission, $n_{\mathrm{b}}=6$. When $\tilde{g}(k)$ takes whole number values, the gear ratios correspond to them, however, for fractional gear numbers, the gear ratios are linearly interpolated. For any other discrete state or control variable, an analogous relaxation will be applied to convert the MINLP into an NLP. Furthermore, since imposing combinatorial constraints (like minimum dwell-time) on discrete variables will not make sense for \textit{relaxed} variables, so we do not impose those in the first step of our three-step algorithm, and take care of it in the second step. Finally, we arrive at the relaxed-NLP (Prob.~\ref{prb:nlp1}) which is solved in step-1 of proposed algorithm by an off-the-shelf gradient-based sparse NLP solver, IPOPT \cite{wachter2006implementation}.
	\begin{problem}[$\mathbb{NLP}\mathnormal{1}$]\label{prb:nlp1}
		It is defined as the discretized equivalent of Prob. \ref{prb:ocp} with a total of $N$ control intervals of step-size $\Delta t=1~\mathrm{s}$ each, using pseudo-spectral collocation scheme at Legendre-Gauss-Radau points for the state trajectories, where:
		\begin{itemize}
			\item a subset of the path constraints $\mathbf{h}\left(\mathbf{x}(t),\mathbf{u}(t),\mathbf{x}_\d(t),\mathbf{u}_\d(t),t\right)\leq 0$, namely the dwell-time constraints (or other combinatorial constraints) on discrete variables are ignored; and
			\item discrete states $\mathbf{x}_\d(t)\in \mathbb{Z}^{|\mathcal{X}_\d|}$ and discrete controls $\mathbf{u}_\d(t)\in \mathbb{Z}^{|\mathcal{U}_\d|}$ are respectively replaced by relaxed states, $\tilde{\mathbf{x}}_\d(t)\in \mathbb{R}^{|\mathcal{X}_\d|}$ and relaxed controls $\tilde{\mathbf{u}}_\d(t)\in \mathbb{R}^{|\mathcal{U}_\d|}$ using linear interpolation.
		\end{itemize} 
	\end{problem}
	\subsection{Constructing the Mixed-Integer Quadratic Program}
	\label{subsec:miqp}
	Assuming that we can obtain a solution, $(\mathbf{x}'(k),\mathbf{u}'(k),\tilde{\mathbf{x}}'_\d(k),\tilde{\mathbf{u}}'_\d(k)) \forall k\in\{1,\cdots,N\}$, to the nonlinear program $\mathbb{NLP}\mathnormal{1}$ we now describe step-2 of our algorithm that handles integer optimization. From the obtained solution, the consistent variables are assigned their fixed trajectories which do not alter after this step:
	\[\mathbf{x}_\mathrm{con}^*(k) \xleftarrow{assign} \mathbf{x}_\mathrm{con}'(k),\qquad \mathbf{u}_\mathrm{con}^*(k) \xleftarrow{assign} \mathbf{u}_\mathrm{con}'(k).\]
	
	Step-1 resulted in relaxed trajectories for the discrete states and controls which are not integer valued nor do they meet dwell-time constraints. Now, the primary focus in step-2 is to obtain the integer states and controls $({\mathbf{x}}_\d^*(k),{\mathbf{u}}_\d^*(k))$ which are \textit{closest} to their relaxed counterparts from step-1, $(\tilde{\mathbf{x}}_\d'(k),\tilde{\mathbf{u}}_\d'(k))$. In doing so, the solution should also satisfy any combinatorial constraints on discrete variables. This is achieved by defining a mixed-integer quadratic program.
	
	\begin{problem}[$\mathbb{MIQP}\mathnormal{2}$]\label{prb:miqp2}
		Given optimized relaxed trajectories of discrete variables $\mathbf{\tilde{x}}'_\d(k)$ and $\mathbf{\tilde{u}}'_\d(k)$, and fixed trajectories of consistent variables $\mathbf{x}_\mathrm{con}^*(k)$ and $\mathbf{u}_\mathrm{con}^*(k)$ for $k\in \{1,\cdots,N\}$ from step-1 solution, solve the corresponding discretized equivalent of the following optimal control problem to obtain $\mathbf{x}_\mathrm{d}^*(k)$ and $\mathbf{u}_\mathrm{d}^*(k)$:
		\begin{align*}
			\min_{\mathbf{u}_\d(t)}~
			&\int_{0}^{T}\left\|\mathbf{u}_\d(t)-\mathbf{\tilde{u}}'_\d(t)\right\|^2 + \left\|\mathbf{x}_\d(t)-\mathbf{\tilde{x}}'_\d(t)\right\|^2\d t
			\\\mathrm{subject~to}\quad
			&\mathbf{\dot{x}}_\d(t)=\mathbf{f}_\d(\mathbf{x}_\d(t),\mathbf{u}_\d(t),t)\\
			&\mathbf{u}_{\d,\l\b} \leq \mathbf{u}_\d(t) \leq \mathbf{u}_{\d,\u\b}\\
			&\mathbf{x}_{\d,\l\b} \leq \mathbf{x}_\d(t) \leq \mathbf{x}_{\d,\u\b}\\
			&\mathbf{x}_\d(0) = \mathbf{x}_{\d,0}\\
			&\mathbf{x}_\d(T) = \mathbf{x}_{\d,T}\\
			&\mathbf{h}\left(\mathbf{x}(t),\mathbf{u}(t),\mathbf{x}_\d(t),\mathbf{u}_\d(t),t\right)\leq 0\qquad\cdots\qquad\\&\qquad\cdots\qquad \begin{cases}\text{for some}~ \mathbf{x}_\mathrm{inc}(t)~\& ~\mathbf{u}_\mathrm{inc}(t)\\\text{satisfying (\ref{eq:stateDyn})(\ref{eq:boundCon})(\ref{eq:boxCon}), where}\\
				\mathbf{u}(t):=\left[\mathbf{u}_\mathrm{inc}(t)~~\mathbf{u}_\mathrm{con}^*(t)\right]^\top\\
				\mathbf{x}(t):=\left[\mathbf{x}_\mathrm{inc}(t)~~\mathbf{x}_\mathrm{con}^*(t)\right]^\top
			\end{cases}
			%\\\mathbf{h}_T\left(\mathbf{x}(T),\mathbf{x}_\d(T),T\right)\leq 0
		\end{align*}
		Note that, since this problem optimizes only the discrete variables, direct shooting discretization scheme is adopted instead of direct pseudo-spectral collocation.
	\end{problem}
	
	Typically, Prob. \ref{prb:miqp2} can be easily framed as a mixed-integer quadratic program. This is because, firstly, the discrete state dynamics, $\mathbf{f}_\d$, are typically linear combinations of shifted and scaled Dirac Delta functions dependent on the integer-valued control | which makes them linear equality constraints. Secondly, for the path constraints $\mathbf{h}\left(\mathbf{x}(t),\mathbf{u}(t),\mathbf{x}_\d(t),\mathbf{u}_\d(t),t\right)$, the consistent variables are already known and fixed. Whereas, only the existence condition of a feasible solution is required for the inconsistent variables thereby avoiding explicit inclusion of nonlinearities in the Prob. \ref{prb:miqp2}. The remaining terms are either quadratic on linear. In the following paragraph, we give an example of Prob. \ref{prb:miqp2} having gear number as the optimization variable with minimum dwell-time and other combinatorial constraints, which is written as an MIQP.
	
	\subsubsection*{Gear Example of $\mathbb{MIQP}\mathnormal{2}$}
	%\label{subsec:MIQPex}
	To better explain Prob. \ref{prb:miqp2}, we given an example MIQP problem that has gear number as a discrete control variable and is imposed with dwell-time constraints. Let's denote the optimal relaxed gear number obtained from step-1 as $\tilde{g}'(k)$. First, we transform the scalar relaxed gear trajectory, $\tilde{g}'(k) \in [1,n_{\mathrm{b}}]\subset \mathbb{R}$ for $k=1,\cdots,N$, into a vectorized binary equivalent, $\mathbf{r}'(k) = \begin{bmatrix}
		r_{1}'(k)	&	r_{2}'(k)	&	\cdots	&	r_{n_{\mathrm{b}}}'(k)
	\end{bmatrix}^\top\in [0,1]^{n_\mathrm{b}}\subset \mathbb{R}^{n_\mathrm{b}}$ where each element of the vector represents one of the $n_{\mathrm{b}}$ gear choices. Representing dwell-time constraint using binary variables that take values $0$ or $1$ is simpler than representing it using integer variables. To arrive at $\mathbf{r}'(k)$ from $\tilde{g}'(k)$, we distribute the percentage difference in between floor $\lfloor \tilde{g}'(k) \rfloor$ and ceil $\lceil \tilde{g}'(k) \rceil$ integers indicating likelihood of belonging to one of the two nearest integer gear numbers. For example, if the relaxed gear took a value of $\tilde{g}'(k) = 3.38$ at $k$-th control interval, then its best integer value has 38\% likelihood of being in 4$^\mathrm{th}$ gear and 62\% of being in 3$^\mathrm{rd}$ gear. The equivalent binary vector for 6-speed transmission will be $\mathbf{r}'(k) = \begin{bmatrix}
		0	&	0	&	0.62	&	0.38	&	0	&	0
	\end{bmatrix}^\top$. Given the relaxed vectorized gear trajectory as an input, the following mixed-integer quadratic program (MIQP) aims to determine the binary-valued vectorized gear trajectory, $\mathbf{b}(k)\in \{0,1\}^{n_\mathrm{b}}~\forall k=1,\cdots,N$. The minimum gear dwell-time constraint needs to be properly defined in discrete-time at this point, and by the use of binary variables, this task is simplifed to two sets of inequalities for each possible gear choice at each time step, as given in Prob.~\ref{prb:miqp2Example}.
	\begin{problem}\label{prb:miqp2Example}
		For every $k\in\left\{1,\cdots,N\right\}$ grid interval and $n_{\mathrm{b}}$ gear choices at each step, obtain the binary gear trajectory, $\mathbf{b}(k)\in \{0,1\}^{n_\mathrm{b}}$ that minimizes sum of its squared differences from the input relaxed gear trajectory, $\mathbf{r}'(k)\in [0,1]^{n_\mathrm{b}}$,
		\begin{align*}
			\min_{\mathbf{b}(k)}~
			&\sum_{k=1}^{N}\left\|\mathbf{b}(k)-\mathbf{r}'(k)\right\|_2^2 = \sum_{k=1}^{N}\!\sum_{j=1}^{n_{\mathrm{b}}}\left(b_{j}(k)-r_{j}'(k)\right)^2
			\\
			\mathrm{s.t.}~
			&\textit{One-Gear-At-A-Time Constraint~}\forall k:\\
			&
			1 = \sum_{j=1}^{n_{\mathrm{b}}} b_{j}(k),
			\\
			&\textit{Feasible Gear Selection Constraint~}\forall k~\forall j:\\
			&
			0\leq b_{j}(k) \leq B_{j}(k) := 
			\begin{cases}
				1\mathrm{~if~}j\text{-}\mathrm{th~gear~is~feasible~at~}k,\\
				0\mathrm{~if~}j\text{-}\mathrm{th~gear~is~infeasible~at~}k.
			\end{cases}\\
			&\textit{Minimum Dwell-Time Constraints~}\forall k~\forall j:\\
			&\forall i\in \left\{k,k+1,\cdots,k+\left\lfloor\frac{t_\mathrm{dwell}}{\Delta t}\right\rfloor\right\}:\\
			&~\qquad b_{j}(k) - b_{j}(k-1) \leq b_{j}(i)\\
			&~\qquad b_{j}(k-1) - b_{j}(k) \leq 1 - b_{j}(i)
		\end{align*}
		where, $t_\mathrm{dwell}$ is the minimum dwell-time duration in seconds that gear has to remain unchanged before next gear shift. Here, gear feasibility limit $B_{j}(k)~\forall k~\forall j$ is pre-calculated using min/max shaft speed limits and torque limits of internal combustion engine and traction motor, based on Prob.~\ref{prb:nlp1}'s solution $(\mathbf{x}_\mathrm{con}^*(k),\mathbf{u}_\mathrm{con}^*(k))$ of consistent variables.
	\end{problem}
	
	As a result of solving Prob.~\ref{prb:miqp2Example}, we obtain the optimal binary-valued discrete variable trajectories $\mathbf{b}(k)$ for $k=1,\cdots,N$. This is transformed back into optimal integer trajectories of the discrete variables $\mathbf{x}_\d^*(k)$ and $\mathbf{u}_\d^*(k)$, which completes integer optimization.
	
	%We note that, as indicated in literature \cite{burger2019design}, Prob.~\ref{prb:miqp2Example} can also be replaced by a mixed-integer linear program by replacing the cost function by a linear function of the decision variables $b_{j}(k)$ while all constraints are already linear in $b_{j}(k)$. We suffice on a MIQP, using $\ell_2$-norm as that is a qualitatively better way of defining the objective, and the mixed-integer programming solver we use has the capability to deal with MIQPs.
	\begin{problem}[$\mathbb{NLP}\mathnormal{3}$]\label{prb:nlp3}
		This is a simpler version of Prob. \ref{prb:nlp1} where the consistent variables $(\mathbf{x}_\mathrm{con}^*,\mathbf{u}_\mathrm{con}^*)$ and discrete variables $(\mathbf{x}_\d^*,\mathbf{u}_\d^*)$ are fixed and known beforehand from steps 1 and 2. The objective is to minimize the fuel cost using only the inconsistent variables $(\mathbf{x}_\mathrm{inc}(k),\mathbf{u}_\mathrm{inc}(k)) \forall k\in\{1,\cdots,N\}$ subject to relevant set of dynamical, box, path and boundary constraints from Prob.~\ref{prb:ocp}.
		%	\begin{align*}
		%	\min_{\zeta_i,m_{\f,k},T_{\b,k},\mu_i}\quad
		%	&\sum_{k=1}^{N}\dot{m}_\f(t) \,\d t&&
		%	\\
		%	\mathrm{subject~to}\quad
		%	&\textit{Multiphase ODEs}\\
		%	&\textit{Box Constraints}\\
		%	&\textit{Path Constraints}\\
		%	&\textit{Boundary Constraints}
		%	\end{align*}
		Here also, there are $N$ control intervals of step-size $\Delta t=1~\mathrm{s}$ each, employing pseudo-spectral collocation using Legendre-Gauss-Radau points.
	\end{problem}
	\subsection{Overall Algorithm}
	\label{subsec:algroithm}
	Referring back to the algorithm diagram of Fig.~\ref{fig:algo_diag}, once $\mathbb{NLP}\mathnormal{1}$ and $\mathbb{MIQP}\mathnormal{2}$ are solved, we have the optimal consistent variables $(\mathbf{x}_\mathrm{con}^*,\mathbf{u}_\mathrm{con}^*)$ (cf. step-1) and the optimal integer variables $(\mathbf{x}_\d^*,\mathbf{u}_\d^*)$ (cf. step-2). These are used to obtain overall optimal solution for the remaining variables, i.e., inconsistent variables $(\mathbf{x}_\mathrm{inc}^*,\mathbf{u}_\mathrm{inc}^*)$ using the nonlinear program defined by Prob.~\ref{prb:nlp3}. This complete process is explained in an algorithmic form in Algorithm \ref{algo}, where writing the parenthesized ``$(k)$'' is avoided for brevity.
	\begin{algorithm}
		\caption{(PS3) Mixed-integer powertrain control using nonlinear programming and pseudo-spectral collocation}
		\label{algo}
		\begin{algorithmic}[1]
			\State{Load drive cycle information, model parameters, and maps}
			\State{Obtain na\"ive initial guess for all the state and control variables $(\bar{\mathbf{x}}_\mathrm{con},\bar{\mathbf{u}}_\mathrm{con},\bar{\mathbf{x}}_\mathrm{inc},\bar{\mathbf{u}}_\mathrm{inc},\bar{\mathbf{x}}_\d,\bar{\mathbf{u}}_\d)$ which can be simply rule-based}
			\State{\textbf{Step 1:} Assume relaxed values $(\tilde{\mathbf{x}}_\d,\tilde{\mathbf{u}}_\d)$ for the integer-valued variables, $(\mathbf{x}_\d,\mathbf{u}_\d)$ and then solve the large nonlinear program (Prob.~\ref{prb:nlp1}) to obtain optimal trajectories of all states $(\mathbf{x}'_\mathrm{con},\mathbf{x}'_\mathrm{inc},\tilde{\mathbf{x}}'_\d)$, and controls $(\mathbf{u}'_\mathrm{con},\mathbf{u}'_\mathrm{inc},\tilde{\mathbf{u}}'_\d)$:
				\[\left(\begin{matrix}\mathbf{x}'_\mathrm{con},\mathbf{x}'_\mathrm{inc},\tilde{\mathbf{x}}'_\d,\\\mathbf{u}'_\mathrm{con},\mathbf{u}'_\mathrm{inc},\tilde{\mathbf{u}}'_\d\end{matrix}\right) \xleftarrow{solve} \mathbb{NLP}\mathnormal{1}\left(\begin{matrix}\bar{\mathbf{x}}_\mathrm{con},\bar{\mathbf{x}}_\mathrm{inc},\bar{\mathbf{x}}_\d,\\\bar{\mathbf{u}}_\mathrm{con},\bar{\mathbf{u}}_\mathrm{inc},\bar{\mathbf{u}}_\d\end{matrix}\right)\] }
			\State{Fix the optimal trajectories of consistent variables from the obtained solution of step-1:
				\[\left(\mathbf{x}_\mathrm{con}^*,\mathbf{u}_\mathrm{con}^*\right) \xleftarrow{assign} \left(\mathbf{x}_\mathrm{con}',\mathbf{u}_\mathrm{con}'\right)\]
			}
			\State{\textbf{Step 2:} Using optimal trajectories of consistent variables $\left(\mathbf{x}_\mathrm{con}^*,\mathbf{u}_\mathrm{con}^*\right)$ and the relaxed variables $\left(\tilde{\mathbf{x}}'_\d,\tilde{\mathbf{u}}'_\d\right)$ solve mixed-integer quadratic program (Prob.~\ref{prb:miqp2}) to obtain integer solutions $(\mathbf{x}_\d^*,\mathbf{u}_\d^*)$ respecting all relevant constraints including the combinatorial constraints. This can be done by transforming integer variables into vectorized binary equivalents:
				\[(\mathbf{x}_\d^*,\mathbf{u}_\d^*) \xleftarrow{solve} \mathbb{MIQP}\mathnormal{2}\left(\mathbf{x}_\mathrm{con}^*,\mathbf{u}_\mathrm{con}^*,\tilde{\mathbf{x}}'_\d,\tilde{\mathbf{u}}'_\d\right) \]}
			\State{\textbf{Step 3:} By fixing the optimal trajectories of discrete variables from step-2 $\left(\mathbf{x}_\d^*,\mathbf{u}_\d^*\right)$ and consistent variables from step-1 $\left(\mathbf{x}_\mathrm{con}^*,\mathbf{u}_\mathrm{con}^*\right)$, solve the second nonlinear program (Prob.~\ref{prb:nlp3}) with step-1's solution as an initial guess. Here, all the inconsistent state and control variables will be re-optimized.
				\[\left(\mathbf{x}^*_\mathrm{inc},\mathbf{u}^*_\mathrm{inc}\right) \xleftarrow{solve} \mathbb{NLP}\mathnormal{3}\left(\mathbf{x}'_\mathrm{inc},\mathbf{u}'_\mathrm{inc}\right)\] }
			\State{The overall mixed-integer solution is finally thus obtained $\left(\mathbf{x}^*, \mathbf{u}^*,\mathbf{x}_\d^*,\mathbf{u}_\d^*\right)$, where the continuous variables are $\mathbf{u}^*=\left[\mathbf{u}_\mathrm{inc}^*~\mathbf{u}_\mathrm{con}^*\right]^\top$ and $\mathbf{x}^*=\left[\mathbf{x}_\mathrm{inc}^*~\mathbf{x}_\mathrm{con}^*\right]^\top$.}
		\end{algorithmic}
	\end{algorithm}
	
	\section{Performance Analysis: PS3 vs. DP}
	\label{sec:results}
	In this section, we have present the usefulness of our proposed algorithm in comparison to the standard benchmark of Dynamic Programming (DP) for HEV energy management problems. We ran experiments for a variety of problems i.e. different combinations of real and integer-valued state and control variables in the OCP, using the algorithms PS3 and DP. The conclusions are summarized in the end of this section. In the sequel \cite{paper2citation}, our main case-study problem which DP is unable to solve is presented in detail, which combines all modeled states and controls into a large mixed-integer optimal control problem. But for the results in this paper, we focus only on five case problems (Case 1 to Case 5) that DP can realistically solve to establish a comparison. In all five cases presented here, all the real-valued variables are inconsistent variables. However, the case-study problem in \cite{paper2citation} involves consistent as well as inconsistent real-valued variables.
	
	All computations were conducted on Lenovo ThinkPad X1 Carbon Laptop PC with an Intel(R) Core(TM) i5-8250U CPU and 8 GB RAM running Windows 10. To model and solve the nonlinear program we use CasADi 3.4.5, \cite{Andersson2019}, within MATLAB R2019a. For easier implementation, a CasADi-based toolbox, YOP \cite{leek2016optimal} is used to parse the optimal control problems into nonlinear programs. The solution to NLP is provided by the sparse NLP solver IPOPT by \cite{wachter2006implementation}, running the linear solver MUMPS or MA97 by Harwell Subroutine Library \cite{hsl2007collection}. For solving MIQP, we use Gurobi optimizer \cite{gurobi}.
	
	For consistent comparison, discretization step-size of one second is chosen for both algorithms (PS3 and DP) and a first-order polynomial degree is used for collocation in PS3 (with LGR points). The initial guess for the optimization variables used by PS3 was based on na\"ive rule-based estimate of state and control trajectories. The DP solver we use is based on the well known `dpm' function method by \cite{sundstrom2009generic}. Space discretizations we used in DP for the state and control variables are provided in Table~\ref{tab:DPdiscretization}.
	\begin{table}[!t]
		\centering
		\caption{DP space discretization levels}
		\label{tab:DPdiscretization}
		\begin{tabular}{|c|c|c|c|c|}
			\hline
			Variable Name [units] & Used in & Space & Lower & Upper\\
			& & Disc. & Bound & Bound\\
			\hline
			Battery State of Charge [-] & Case 1-5 & 61 & 0.3 & 0.8\\
			Torque Split [-] & Case 1-5 & 21 & $-$1 & 1\\
			Battery Temperature [$^{\circ}$C] & \!Case 2\,\&\,4\! & 8 & 23 & 30 \\
			Gear Number [-] & Case 3-4 & 6 & 1 & 6 \\
			Gear Dwell-time Counter [s] & Case 3-4 & 5 & 0 & 4 \\
			Gear Shift [-] & Case 3-4 & 3 & $-$1 & 1 \\
			Vehicle Speed [m/s] & Case 5 & 26 & 0 & 25 \\
			Vehicle Position [m] & Case 5 & 65 & 0 & $\sim$6463 \\
			Vehicle Acceleration [m/s$^2$] & Case 5 & 15 & $-$2 & 1.5 \\
			\hline
		\end{tabular}
	\end{table}
	\subsection{Architecture, Models and Drive Cycle}
\begin{figure}[!t]
		\centering
		\includegraphics*[width=\linewidth]{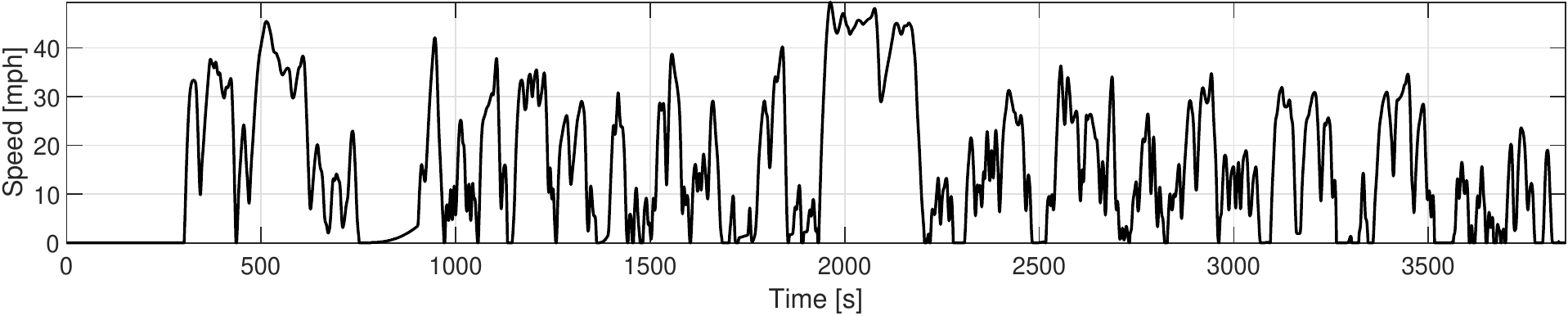}
		\caption{NREL Parcel Delivery Truck Cycle (Baltimore) \cite{nreldrivetool} as reference drive cycle, $v_{\org}$.\label{fig:drivecycle}}
	\end{figure}

\begin{figure*}[!t]
		\centering
		\includegraphics*[width=\textwidth]{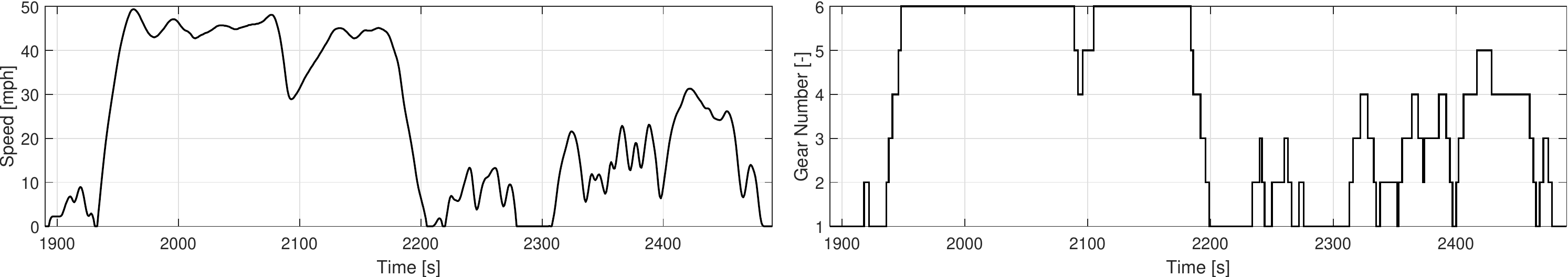}
		\caption{Drive cycle segment (left) and corresponding pre-determined gear profile (right) used for numerical experiments.\label{fig:drivecycle_wgear}}
	\end{figure*}

	For all our case studies in this paper and the main case-study in sequel \cite{paper2citation}, we have considered a strong parallel P2 hybrid electric vehicle architecture, like the one shown in Fig.~\ref{fig:architecture}. This is for a medium-duty diesel-engine, a $90~\mathrm{kW}$-rated electric machine, $11~\mathrm{kWh}$ lithium-iron-phosphate (LFP) battery pack with $31~\mathrm{Ah}$ capacity, and 6-speed automatic transmission. Data maps for the LFP battery model used in this paper are given in the \ref{app:hevmodel}. All other modeling details about the internal combustion engine, electric machine, vehicle dynamics and driveline are given in the sequel \cite{paper2citation}.
	%Note that, even though a specified speed profile is provided as an input, the controller determines what speed profile to actually follow i.e. the eco-driving speed profile. The input speed profile, $v_{\org}(t)$ only serves as a reference trajectory using which certain constraints are defined in the problem formulation.
	A short segment of 10-minutes duration from the NREL drive cycle for parcel pick-up and delivery, Fig.~\ref{fig:drivecycle}, with a speed-dependent gear profile is used as reference for the following experiments, as shown in Fig.~\ref{fig:drivecycle_wgear}.
	\subsection{Case 1: Basic Hybrid (1S1C)}
	\label{subsec:basicproblem}
	This problem involves a single real-valued state, battery state-of-charge, SOC ($\zeta$) and a single control variable, torque split between internal combustion engine and the electric motor ($\mu$). Since no integer variables are involved in this problem, hence, only the step-1 of Algorithm \ref{algo} is relevant and used which gives the final optimal solution. 
	
	The control and state-space discretization required for DP is set to take 61 values for SOC ($0.3\leq \zeta \leq 0.8$), and 21 values for the control variable, torque split ($-1\leq \mu \leq 1$). This discretization is chosen to keep minimum computation time and memory load, without significant drop in optimality of the solutions. A point to note is that unlike DP, PS3 can take all real-values up-to machine precision for the state and control variables, i.e., its search space is not discretized the way it is for DP.
	
\begin{figure}[!t]
		\centering
		\includegraphics*[width=\linewidth]{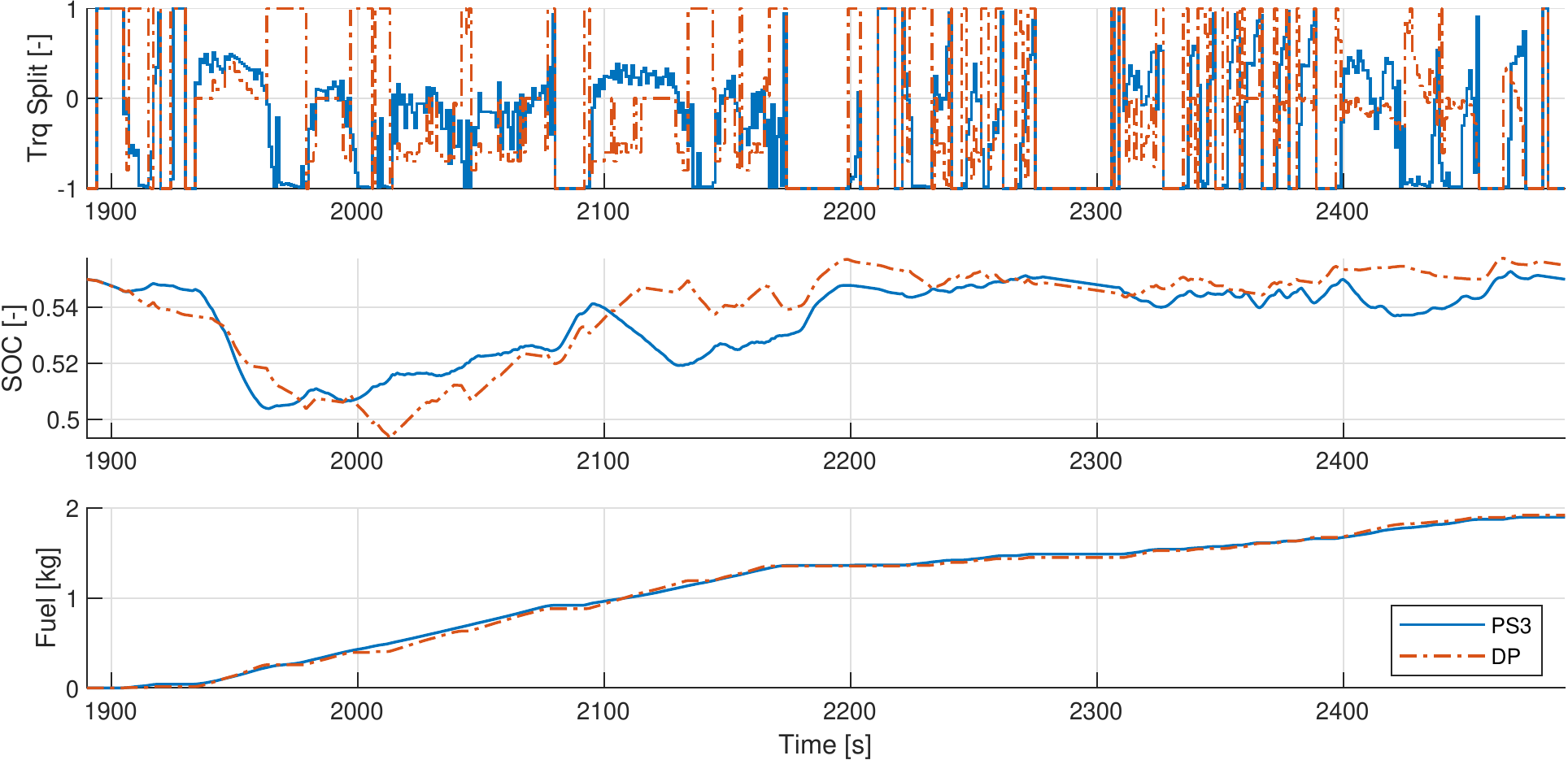}
		\caption{PS3 versus DP: comparable results for basic hybrid 1S1C problem.\label{fig:basicproblem_result}}
	\end{figure}

	The obtained results are plotted in Fig.~\ref{fig:basicproblem_result}. Although the state (SOC) and control (Torque Split) trajectories appear different at many places in the plot, we observe that both the algorithms have comparable overall cost i.e. total fuel consumed | \textbf{1.920 kg} (DP) and \textbf{1.921 kg} (PS3), and have low computational times | \textbf{4 seconds} (DP) and \textbf{19 seconds} (PS3).
	
	\subsection{Case 2: Thermal Hybrid (2S1C)}
	\label{subsec:thermalhybrid}
\begin{figure}[!t]
		\centering
		\includegraphics*[width=\linewidth]{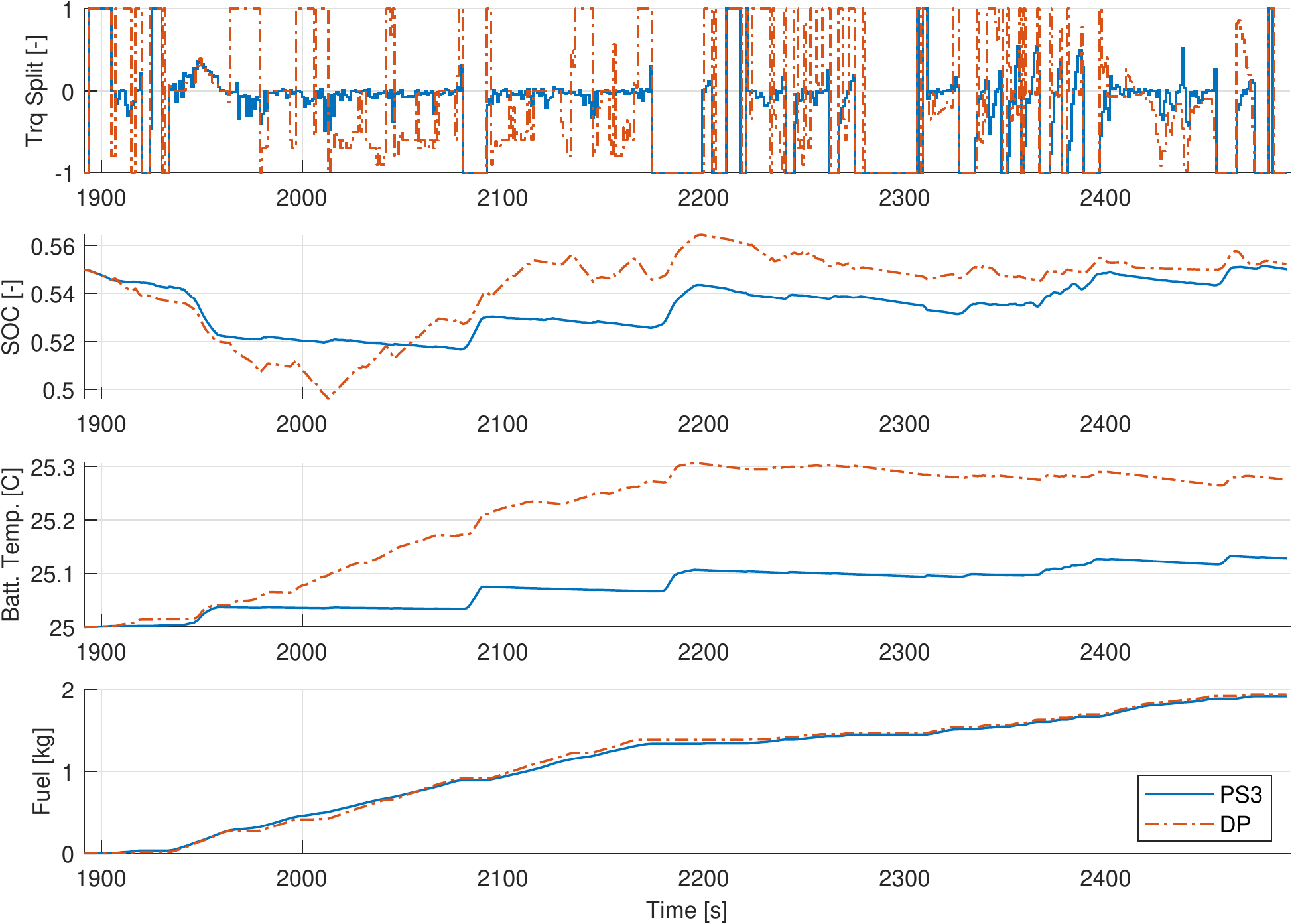}
		\caption{PS3 versus DP: comparable results for thermal hybrid 2S1C problem.\label{fig:case2result}}
	\end{figure}

	The second problem builds on top of the basic hybrid problem by involving two real-valued state variables, battery SOC and battery temperature, and one control variable, torque split. With the additional state variable of battery temperature having first-order thermal dynamics model, we make use of temperature-dependent (and SOC-dependent) 2D look-up table for cell internal resistance (see \ref{app:hevmodel} for its modeling details). Our LFP battery model has very low Ohmic heat loss for a 10-minute drive cycle. In fact, the overall change in battery temperature is within one Celsius of the ambient temperature (25 Celsius). For this reason, we discretize the battery temperature values in DP to take any of 8 uniformly-spaced values within 23 C and 30 C. Results are plotted in Fig.~\ref{fig:case2result}. Again, we observe that performance is comparable | \textbf{1.93 kg} (DP) and \textbf{1.91 kg} (PS3), and computational times are still low | \textbf{8.21 seconds} (DP) and \textbf{23.27 seconds} (PS3). Furthermore, the trajectories are different, yet overall effect on the cost is similar. An observation is that DP has higher battery utilization instances causing more current to be drawn in-and-out, and hence, the battery temperature rises more in the DP solution.
	
	\subsection{Case 3: Gear Hybrid (1S1C \& 2D$_\text{S}$1D$_\text{C}$)}
	\label{subsec:gearhybrid}
\begin{figure*}[!t]
		\centering
		\includegraphics*[width=\textwidth]{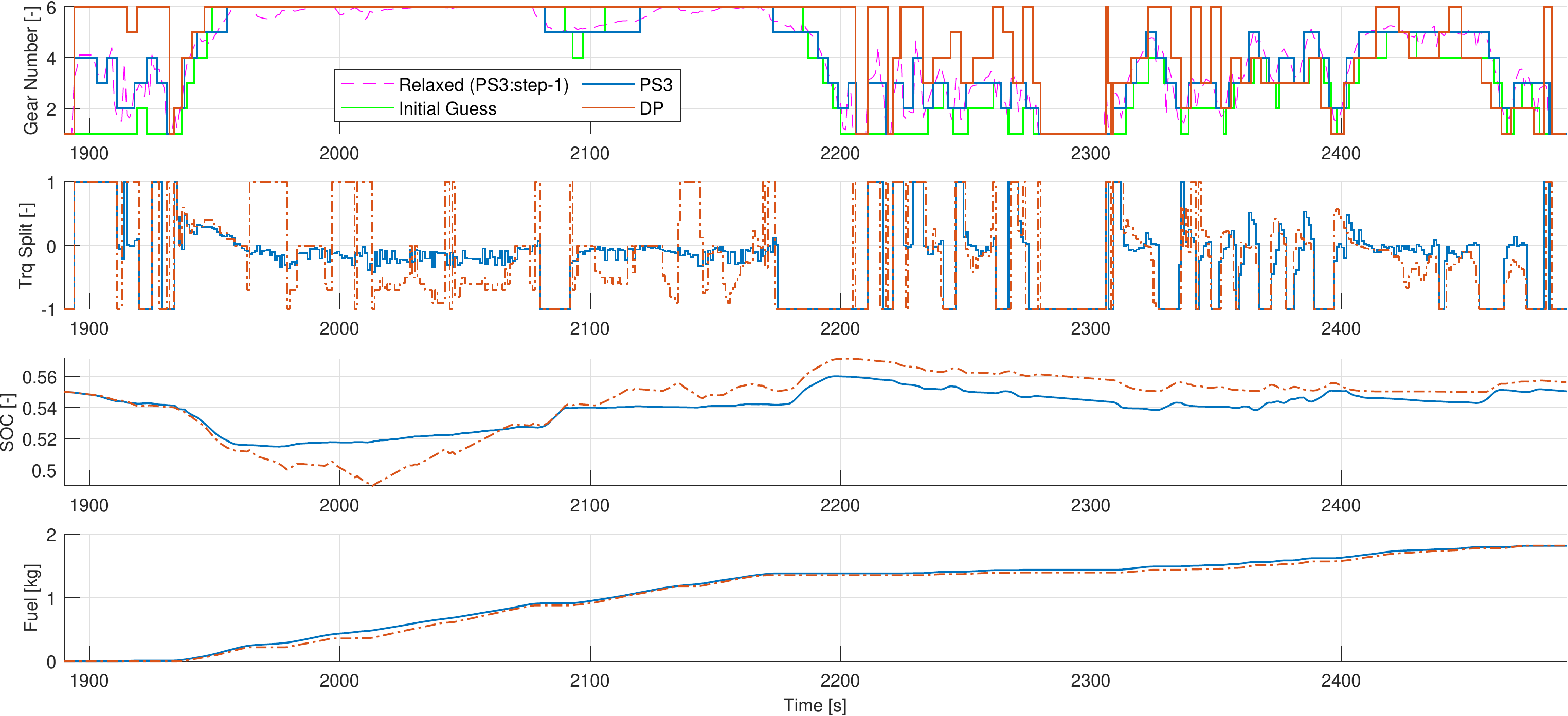}
		\caption{PS3 versus DP: results for gear hybrid (1S1C \& 2D$_\text{S}$1D$_\text{C}$) problem.\label{fig:case3result}}
	\end{figure*}

	This case involves a mixed-integer optimal control problem. It considers one real-valued state, battery SOC, and one control variable, torque split (1S1C). And there are two integer-valued states, gear number and gear dwell-time counter, and one integer-valued control, gear shift command (2D$_\text{S}$1D$_\text{C}$). As for DP, the space-discretization for real-valued variables is the same as before, and the integer-valued variables have search space at only their respective feasible integer values (e.g. gear number can be an integer from $1$ to $6$, gear command can be an integer from $-5$ to $5$, etc.). Being a mixed-integer problem, we make full use of the three-step algorithm, PS3. In the first step, we obtain a relaxed gear profile (shown in the results plot later). Second step solves a mixed-integer quadratic program to find an integer gear profile \textit{near} the relaxed profile while meeting the 3-seconds dwell-time constraint. And finally, the third step obtains the optimal real-valued signals with the input of known gear profile obtained from second step. Some important signals are shown in the plots of Fig.~\ref{fig:case3result}.
	Key observations for this problem from the plots are:
	\begin{itemize}
		\item Gear profiles of PS3 (blue) and DP (orange) are quite different, and so are the torque split profiles. But, the total fuel consumed by the end is almost identical | \textbf{1.819 kg} (DP) and \textbf{1.815 kg} (PS3).
		\item Computational times are \textbf{502.82 seconds} (DP) and \textit{[1249.9+1.4786+16.1430=]} \textbf{1267.5 seconds} (PS3). DP's computational load is still tractable because the most of the state or control variables involved are integer-valued in this problem. DP is able to handle integer-valued variables quite well because space-discretization is simplified for integers. Computation time for PS3 is higher, but still tractable, than that for DP as it requires running of three sequential computational programs to be solved over its three steps.
		\item Gear profile from DP solution tends to take higher values, which is better for fuel reduction, but it comes at the expense of steeper drops in the SOC. To meet the charge-sustaining constraint, DP solution then uses larger magnitudes of engine torque values costing higher fuel. The net result is that DP's fuel trajectory is lower in the first half of the cycle, but by the end it meets up with that of PS3 resulting in identical overall fuel consumed.
		\item Since we use the interior-point solver, IPOPT as the NLP solver for PS3, we see that the relaxed gear trajectory (dashed-magenta in first subplot) of PS3's step-1 is close to the given rule-based initial guess (solid-green). This confirms that having a good initial guess, especially when integer-valued variables are involved is critical for optimality in solutions.
	\end{itemize}
	
	\subsection{Case 4: Thermal Gear Hybrid  (2S1C \& 2D$_\text{S}$1D$_\text{C}$)}
	\label{subsec:thermalgearhybrid}
\begin{figure}[!t]
		\centering
		\includegraphics*[width=\linewidth]{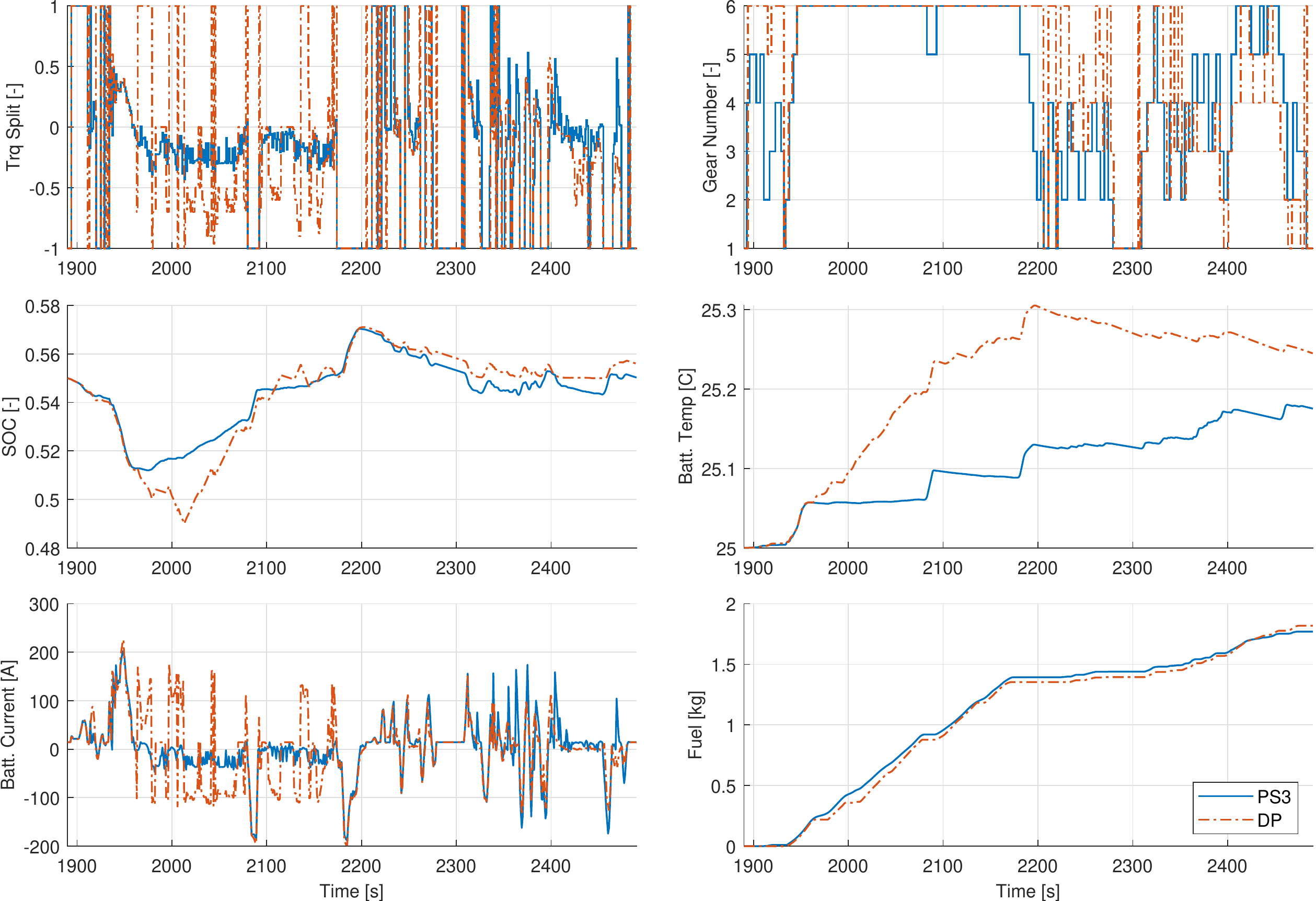}
		\caption{PS3 versus DP: comparable results for thermal gear hybrid (2S1C \& 2D$_\text{S}$1D$_\text{C}$) problem.\label{fig:case4result}}
	\end{figure}

	In order to demonstrate, how DP starts to become intractable for more complicated problems, we consider this case of combined Cases 2-3, to make Case 4. Essentially, on top of the state and control variables of the gear hybrid problem, now we have a fourth state variable, that of battery temperature which is real-valued like SOC. So, this problem involves two real-valued and two integer-valued state variables, and similarly one real-valued and one integer-valued control variable. Some plots for the obtained results are shown in Fig.~\ref{fig:case4result}.
	The key take-away from this experiment is that PS3 remains computationally reasonable despite to an added real-valued variable, however, due to its curse of dimensionality, DP starts to require large memory and computational resources. Overall fuel consumed is \textbf{1.818 kg} (DP) and \textbf{1.770 kg} (PS3), and computational times are \textbf{5046.6 seconds (84 minutes)} (DP) and \textit{[1843.5+1.5770+63.6420=]} \textbf{1908.7 seconds (32 minutes)} (PS3).
	
	\subsection{Case 5: Eco Hybrid (3S2C)}
	\label{subsec:ecohybrid}
\begin{figure*}[!t]
		\centering
		\includegraphics*[width=\textwidth]{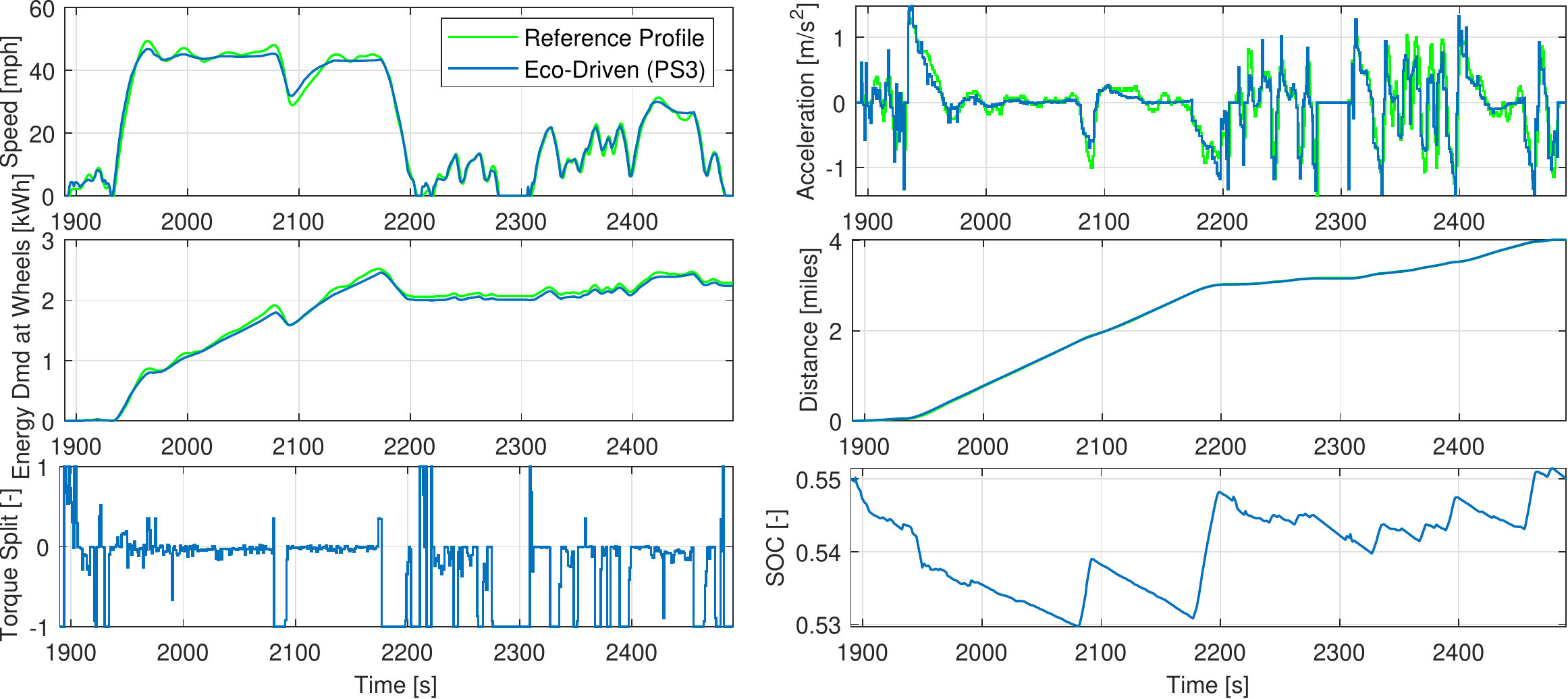}
		\caption{PS3 results for eco hybrid 3S2C problem.\label{fig:case5result}}
	\end{figure*}

	Finally, we consider a case in which we have three state variables (SOC, vehicle speed, vehicle position) and two control variables (SOC, vehicle acceleration). The idea of eco-driving is to allow the vehicle to maneuver within a 5 km/h threshold of the reference target speed profile (shown in green in third subplot of Fig.~\ref{fig:case5result}), such that the total distance covered on the whole route is the same. Simultaneously, torque split is also optimized.
	
	PS3 is able to give a solution which is shown in Fig.~\ref{fig:case5result}. Computation time running PS3 for this experiment was \textbf{585.62 seconds} and the total fuel consumed was \textbf{1.90 kg}. The benefit of eco-driving versus non-eco-driving scenario was measured by comparing the net energy demand at the wheels which reduces by $2.213\%$ of the reference, shown in the last plot of Fig.~\ref{fig:case5result}.
	
	As for a solution using Dynamic Programming, since all these variables are real-valued, DP exceeds the memory resources and fails to give a solution. When we tried a coarse space-discretization that does not exceed available memory, it is unable to find a feasible solution due to the coarseness. 
	
	When analyzing the plots, we observe that the effect of eco-driving is that the eco-driven vehicle operates at slightly lower speeds when the reference is at high speeds, and at slightly higher speed when the reference is at very low speed | this behavior allows the eco-driven vehicle to use more electrical energy for vehicle traction, instead of fuel energy, thereby reducing fuel.
	
	\subsection{Discussion}
	In Table~\ref{tab:summarytable} we have summarized the total fuel consumed for the various problems we present to benchmark performance of PS3 against DP. We observe that for problems which DP can solve, PS3's solutions match DP's globally optimal solutions. This establishes the general acceptability of PS3 as an alternative benchmark against DP. We do note that theoretically, DP is a global optimization solver, and PS3 | using IPOPT | only gives locally optimal solutions. But, as we saw in results of all practical case studies, robust adjustment of solver parameters and initial guesses leads PS3 to highly useful and near-globally-optimal performance, within reasonable computational cost.
	\begin{table}[!t]
		\begin{center}
			\caption{PS3 versus DP: Fuel consumed [kg] for various problems. Here, [D] refers to an integer variable\label{tab:summarytable}}
			%\begin{tabular}{ | m{5cm} | m{5cm}| m{5cm} | m{5cm} | m{5cm} |}
			\begin{tabular}{  m{0.01\linewidth} | m{0.17\linewidth} | m{0.25\linewidth} | m{0.17\linewidth} | m{0.05\linewidth}| m{0.05\linewidth}  }
				%\hline
				& \textbf{Case} & \textbf{States} & \textbf{Controls} & \textbf{DP} & \textbf{PS3} \\
				\hline
				\hline
				1& Basic Hybrid (1S1C) & SOC & Torque Split &  1.92 & 1.92 \\
				\hline
				2& Thermal Hybrid (2S1C) & SOC, Battery Temperature & Torque Split & 1.93 & 1.91 \\
				\hline
				3& Gear Hybrid (1S1C \& 2D$_\text{S}$1D$_\text{C}$) & SOC, Gear Number [D], Gear Dwell Time [D] & Torque Split, Gear Shift [D] & 1.82 & 1.82 \\
				\hline
				4& Thermal Gear Hybrid (2S1C \& 2D$_\text{S}$1D$_\text{C}$) & SOC, Battery Temperature, Gear Number [D], Gear Dwell Time [D] & Torque Split, Gear Shift [D] & 1.82 & 1.77 \\
				\hline
				5& Eco Hybrid (3S2C) & SOC, Vehicle Speed, Vehicle Position & Torque Split, Vehicle Acceleration & N/A & 1.90 \\
				%\hline
			\end{tabular}
		\end{center}
	\end{table}
	
\begin{figure}[!t]
		\centering
		\includegraphics*[width=\linewidth]{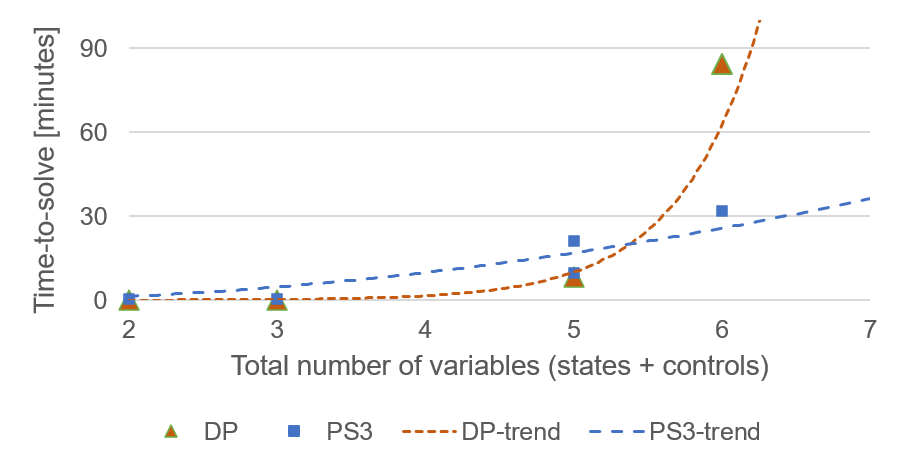}
		\caption{PS3 versus DP: estimated computation time comparison chart.\label{fig:computationComp}}
	\end{figure}

	Secondly, despite PS3 utilizing gradient-based optimization approach, the integer-valued variables (Case 3-4) are well optimized giving benchmark performance with the help of state-of-art MIQP solver, Gurobi. Thirdly, it is a known fact that as the number of state and control variables in an optimal control problem increase, it becomes tedious for dynamic programming to give tractable solutions without compromising sensible space-discretization levels. This is due to DP's inherent curse of dimensionality, and is particularly apparent when real-valued variables are involved. Based on the presented results in this paper, we give an estimate of comparative trends of computation time with increasing problem size in Fig.~\ref{fig:computationComp}. The extrapolated trend of PS3's computation time (dashed-blue line) is backed up with numerical results on a case-study problem involving 13 state variables and 4 control variables, which is presented in the sequel \cite{paper2citation}. Lastly, we note that as with any numerical solver, the computation time for PS3 can vary based on the desired tolerances, initial guess, and other solver options.
	
	\section{Conclusion}
	\label{sec:conclusion}
	In this paper, we have presented a novel algorithm, namely ``PS3'', for mixed-integer optimal control problems with application to energy management of electrified powertrains involving high number of states and controls. It employs direct pseudo-spectral collocation for highly accurate state dynamics estimation and relies on state-of-art numerical optimization solvers for NLP and MIQP. The underlying framework is built upon the open-source modeling language CasADi \cite{Andersson2019}, is implemented in MATLAB, utilizes YOP \cite{leek2016optimal} for parsing NLPs, and runs IPOPT \cite{wachter2006implementation} and Gurobi \cite{gurobi} solvers in its three steps.
	
	%In literature, none of the other state-of-the-art approaches for HEV energy management have addressed simultaneous optimization of electrical (SOC), vehicular (eco-driving) and thermal (battery temperature) dynamics along with an integer (gear) control and its corresponding constraints (dwell-time).
	Our algorithm utilizes validated powertrain component models and stands out in being able to provide solutions to diverse class of powertrain problems. PS3 benchmarks for problems that may involve simultaneous eco-driving and integer optimization along with non-differentiable look-up tables, thermal states, and combinatorial path constraints in the models. We provide empirical justification of PS3's ability to be considered a benchmark algorithm by comparing results against Dynamic Programming for four out of five case-studies where various combinations of continuous and discrete states and controls were chosen to minimize fuel. Results were analysed on a realistic drive cycle with frequent starts and stops, steeper acceleration and deceleration events, as well as wide-range of power demands. Our analysis shows that this algorithm does not scale in computational load as DP does, and can handle highly complex interactions that occur in modern-day powertrains. This methodology can be robustly applied to difficult real-world problem classes and it has potential applications in real-time embeddable controllers. It can even serve as a numerical benchmark for other methodologies in future. In the sequal paper \cite{paper2citation}, we demonstrate, analyze, and benchmark a large case-study problem to show the great benefit of large-scale optimization this method offers.
	
	\appendix
	\section{Battery Model}
	\label{app:hevmodel}
\begin{figure}[!t]
		\centering
		\includegraphics[width=0.49\linewidth]{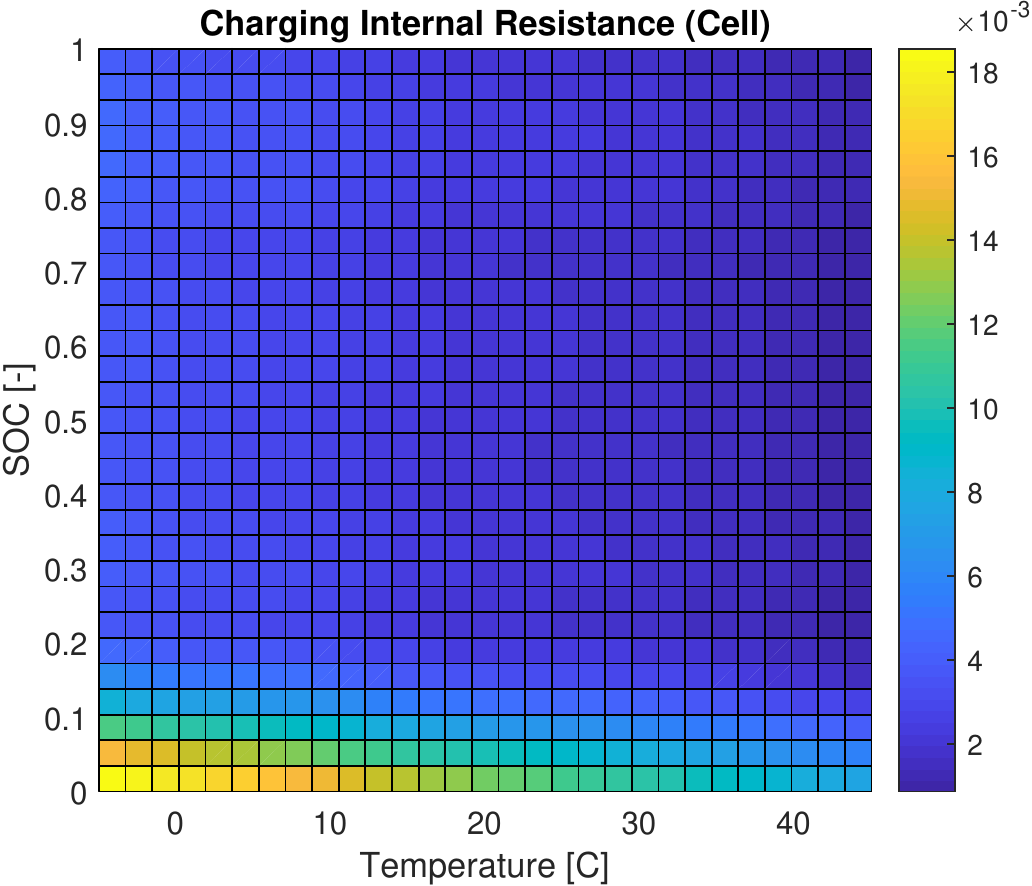}
		\includegraphics[width=0.49\linewidth]{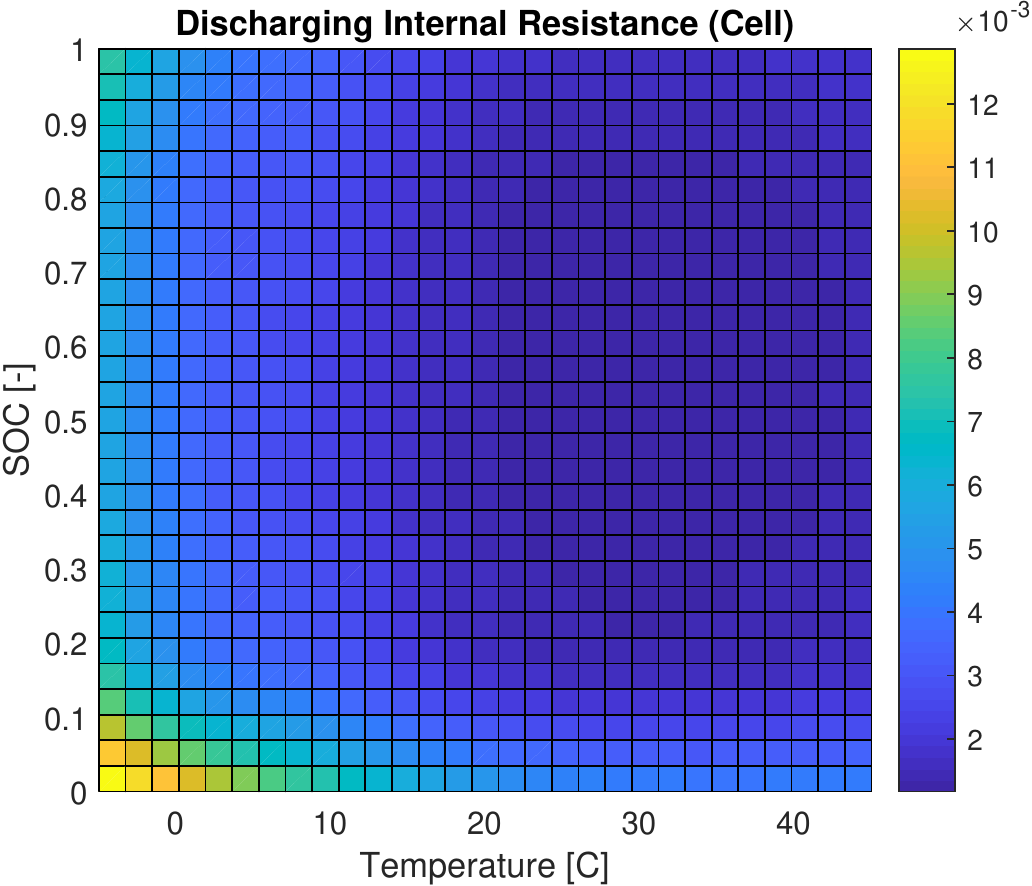}
		\includegraphics[width=0.6\linewidth]{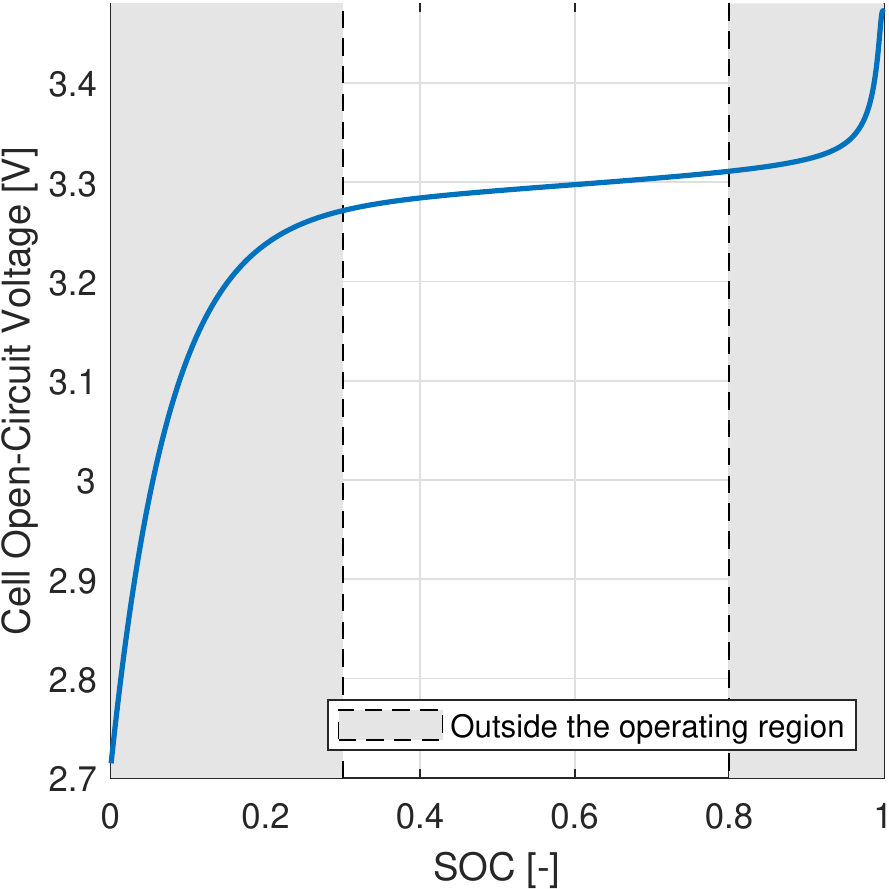}
		\caption{Resistance (left: charging, right: discharging) in Ohms, open-circuit voltage (bottom)}
		\label{fig:battery_R0}
	\end{figure}

	In the appendix, the battery model is presented that is used for the shown experiments and numerical results for the hybrid electric powertrain. The battery pack model used is of 11 kWh Lithium-Iron-Phosphate (LFP) having 350 V nominal voltage. Charge sustaining operation is assumed for the drive cycle, and so the initial condition and final condition for SOC is set equal to 55\%. For the electrical dynamics, we assume a zero-th order equivalent circuit model, and for the thermal dynamics, a first order temperature model with heat addition due to ohmic losses. The equations with further details are given in the sequel \cite{paper2citation}. However, the only difference between the battery model presented in \cite{paper2citation} and this is that here we have temperature-dependent internal resistance 2-D maps, which, along with the open-circuit voltage (OCV) plot is shown in Fig.~\ref{fig:battery_R0}. OCV is modeled using the following expression:
	\[V_{\mathrm{oc}} \!= \!N_\mathrm{s} \!\left( V_0 + \alpha_\b \left( 1 - e^{- \beta_\b \zeta } \right) +\! \gamma_\b \zeta \!+ \zeta_\b \left( 1 - e^{ - \epsilon_\b / ( 1 - \zeta ) } \right)\!\right)\!,\]
	where, $\zeta, N_\mathrm{s}$ and $V_0$ are battery state-of-charge (SOC), number of cells in series, and nominal voltage respectively, while the remaining constants $\alpha_\b, \beta_\b, \gamma_\b, \zeta_\b, \epsilon_\b$ are obtained by curve-fitting the OCV with respect to SOC using real-world empirical data.
	
	\bibliographystyle{IEEEtran} 
	\bibliography{biblibrary}
	%%%%%%%%%%%%%%%%%%%%%%%%%%%%%%%
	
	%\appendices
	
	%\section*{Acknowledgment}
	
\end{document}